\pdfoutput=1
\documentclass[PAPER, atlasdraft=flase, USenglish, cernpreprint, texmf, orcidlogo]{atlasdoc}
\usepackage[subcaption]{atlaspackage}
\usepackage{multirow}
\usepackage{booktabs}
\usepackage{siunitx}
\usepackage{atlasbiblatex}

\usepackage{atlasphysics}
\graphicspath{{logos/}{figures/}}

\usepackage{ANA-EXOT-2018-56-PAPER-defs}
\AtlasTitle{Search for soft unclustered energy patterns containing muons in the final state in $pp$ collisions at $\sqrt{s}=13$~TeV with the ATLAS detector}

\AtlasAbstract{%
Soft unclustered energy patterns (SUEPs) refer to high-multiplicity, isotropic distributions of low-momentum particles that could arise in strongly-coupled hidden sector theories.
A search for SUEPs whose decay products contain muons in the final state is presented using $140~\mathrm{fb}^{-1}$ of proton--proton collision data at $\sqrt{s}=13$~TeV collected by the ATLAS detector at the Large Hadron Collider during 2015--2018.
The analysis targets SUEP signatures via gluon--gluon fusion production of a scalar mediator in Hidden Valley scenarios.
No significant excess over the Standard Model expectation is observed.
Exclusion limits are set on the product of the mediator production cross section and the branching fraction for its decay into a SUEP, down to 0.05~fb for a mediator mass of $m_S=750$~GeV, 0.4~fb for $m_S=400$~GeV, and 70~fb for $m_S=125$~GeV;
the last value, when identifying the mediator with the Standard Model Higgs boson, translates to an upper limit on the branching fraction of around 0.2\%.
}

\PreprintIdNumber{CERN-EP-2026-148}

\AtlasJournal{JHEP}
\hypersetup{pdftitle={ATLAS document},pdfauthor={The ATLAS Collaboration}}

\begin{document}

\maketitle

\tableofcontents

\section{Introduction}
\label{sec:intro}

Hidden Valley (HV) models~\cite{Strassler:2006ri,Strassler:2006im} provide a well-motivated extension of the Standard Model (SM) in which a hidden sector, neutral under SM gauge interactions, is connected to the SM visible sector through weak, so-called \textit{portal} interactions.
Such scenarios arise naturally in solutions to the hierarchy problem, which often introduce additional sectors with suppressed couplings to the SM~\cite{Chacko:2005pe}.
Strongly interacting hidden sectors are further motivated by the possibility that dark matter is composed of bound states of a QCD-like hidden force~\cite{Kaplan:2009ag, Bai:2013xga}, or that it resides in a dark sector with sizable self-interactions,
as suggested by small-scale structure anomalies in galaxies and their clusters~\cite{Wandelt:2000ad,Rocha:2012jg,Tulin:2017ara}.

At the Large Hadron Collider (LHC), a range of searches has targeted these scenarios in regimes where the hidden sector undergoes a shower analogous to a QCD parton shower, giving rise to jet-like signatures once the hidden-sector particles decay to visible SM particles.
These include searches for so-called semi-visible jets~\cite{CMS:2021dzg,ATLAS:2023swa,ATLAS:2025kuz} and emerging jets~\cite{CMS:2024gxp,ATLAS:2025bsz}, which exploit the resulting exotic jet topologies to distinguish such events from SM backgrounds.

\begin{figure}
\centering
\includegraphics[width=0.5\linewidth]{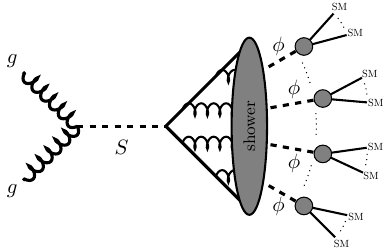}
\caption{
Representative diagram of gluon-fusion production of a scalar mediator \(S\), followed by showering in the dark sector and the production of a high-multiplicity final state of light dark (pseudo-)scalars (\(\phi\)).
The \(\phi\) particles subsequently decay promptly to SM particles through their coupling to dark photons, which mix kinetically with the SM photon.
The figure is taken from Ref.~\cite{Knapen:2016hky}.
}
\label{fig_signal_diagram}
\end{figure}

However, if the hidden-sector force remains strongly coupled well above its hadronization scale, corresponding to a quasi-conformal coupling regime, its radiation pattern can differ qualitatively from that of QCD.
In particular, emissions need not be dominated by soft and collinear splittings, and the resulting shower may instead populate phase space more uniformly~\cite{Strassler:2008bv,Polchinski:2002jw,Lin:2007fa,Hofman:2008ar}.
For models in which the available energy is distributed among many particles, this results in a characteristic collider signature known as a \textit{Soft Unclustered Energy Pattern} (SUEP),
consisting of a high multiplicity of low-momentum particles, distributed approximately isotropically in the center-of-mass frame~\cite{Strassler:2008bv,Knapen:2016hky}.
Figure~\ref{fig_signal_diagram} shows a representative signal diagram. The absence of hard or collimated objects makes such signatures challenging both to trigger on and to reconstruct.
Exploring such unconventional and experimentally challenging signatures is essential to ensure sensitivity to a broad range of possible extensions of the SM.

Both the existing searches for SUEP signatures performed by the CMS Collaboration~\cite{CMS:2024nca,CMS:2026jia}
and the analysis presented here are based on the theoretical framework proposed in Ref.~\cite{Knapen:2016hky}.
In this framework, the hidden sector is accessed through a scalar portal, which may be the SM Higgs boson or a heavier scalar particle.
Within the dark sector, particles undergo showering and subsequently hadronize into a large number of dark mesons \(\phi\), which are assumed for simplicity to consist of a single particle species.
A wide range of values for the dark-meson mass \(\mphi\) is theoretically motivated.
Generally, values of order 1--10~GeV are well suited for SUEP scenarios,
as they allow for a rich set of decay channels to SM particles while still producing
the high-multiplicity, soft final states characteristic of this signature, which are not efficiently targeted by conventional searches.
The momenta of the dark mesons are expected to follow an approximately thermal distribution~\cite{Hatta:2008qx}
and are modeled using a Maxwell--Boltzmann spectrum parameterized by a Hagedorn temperature \(T\)~\cite{Hagedorn:1965st},
which sets the characteristic momentum scale of the hadronizing hidden sector and is expected to be of the same order as \(\mphi\).
The dark mesons are assumed to decay promptly into pairs of dark photons (\(A'\)),
which subsequently decay promptly into charged SM particles via kinetic mixing with the SM photon~\cite{Graham:2021ggy}.
While other decay modes of dark mesons to SM particles are possible~\cite{Knapen:2021eip},
this choice provides a convenient and realistic framework to realize a wide range of final-state particle compositions.
For a given dark-photon mass, the branching fractions to SM particles are fixed, leading to specific mixtures of electrons, muons, and hadrons in the final state.
For dark-photon masses of order 1~GeV, the dominant decay modes are $A'\rightarrow e^+e^-$, $A'\rightarrow\mu^+\mu^-$, and $A'\rightarrow\pi^+\pi^-$.

A previous search for SUEP signatures by the CMS Collaboration~\cite{CMS:2024nca} targeted gluon--gluon fusion production of the mediator and
relied on high-transverse-momentum initial-state radiation to trigger events via large energy deposits in the calorimeters.
While this approach provided sensitivity over a broad range of $T$ and $\mphi$, it was primarily sensitive to scenarios with large mediator masses.
Subsequent searches targeting associated production of the Higgs boson with a vector boson~\cite{CMS:2026jia} complemented this result.
However, the requirement of very high particle multiplicities to suppress background limited the sensitivity to scenarios with low $\mphi$ and $T$.

The search presented here complements these approaches by targeting final states with a high multiplicity of prompt muons, which are expected for a broad range of dark-photon masses $\mAprime$.
This strategy provides strong background rejection while maintaining sensitivity to scenarios with moderate particle multiplicities,
thereby extending the coverage of the SUEP parameter space well beyond that of previous searches.

Events are selected using multi-muon triggers and target prompt final states with a high multiplicity of muons.
They are further characterized by relatively low average muon transverse momentum, large charged-particle track multiplicity associated with the primary interaction,
and isotropic spatial distribution of the muons.
The signal region is defined to be sensitive to a broad range of charged-particle multiplicities and to preferentially select events in which muons
are distributed more isotropically than in the dominant background from QCD multijet production,
where muons are produced mainly in heavy-flavor decays, and which is estimated using a data-driven method.

\section{ATLAS detector}
\label{sec:detector}

% The next lines are included from the .//atlas-detector.tex input file
\newcommand{\AtlasCoordFootnote}{%
ATLAS uses a right-handed coordinate system with its origin at the nominal interaction point (IP)
in the centre of the detector and the \(z\)-axis along the beam pipe. The \(x\)-axis points from the
IP to the centre of the LHC ring, and the \(y\)-axis points upwards. Polar coordinates \((r,\phi)\)
are used in the transverse plane, \(\phi\) being the azimuthal angle around the \(z\)-axis. The
pseudorapidity is defined in terms of the polar angle \(\theta\) as \(\eta = -\ln \tan(\theta/2)\)
and is equal to the rapidity
\( y = \frac{1}{2} \ln \left( \frac{E + p_z}{E - p_z} \right) \)
in the relativistic limit. Angular distance is measured in units of
\(\Delta R \equiv \sqrt{(\Delta y)^{2} + (\Delta\phi)^{2}}\).}

The ATLAS detector~\cite{PERF-2007-01} at the LHC covers nearly the entire solid angle around the
collision point.\footnote{\AtlasCoordFootnote} It consists of an inner tracking detector surrounded
by a thin superconducting solenoid, electromagnetic and hadronic calorimeters, and a muon
spectrometer incorporating three large superconducting air-core toroidal magnets.

The inner-detector system (ID) is immersed in a \qty{2}{\tesla} axial magnetic field and provides
charged-particle tracking in the range \(|\eta| < 2.5\). The high-granularity silicon pixel detector
covers the vertex region and typically provides four measurements per track, the first hit generally
being in the insertable B-layer (IBL) installed before Run~2~\cite{ATLAS-TDR-19,PIX-2018-001}. It is
followed by the SemiConductor Tracker (SCT), which usually provides eight measurements per track.
These silicon detectors are complemented by the transition radiation tracker (TRT), which enables
radially extended track reconstruction up to \(|\eta| = 2.0\). The TRT also provides electron
identification information based on the fraction of hits (typically 30 in total) above a higher
energy-deposit threshold corresponding to transition radiation.

The calorimeter system covers the pseudorapidity range \(|\eta| < 4.9\). Within the region
\(|\eta| < 3.2\), electromagnetic calorimetry is provided by barrel and endcap high-granularity
lead/liquid-argon (LAr) calorimeters, with an additional thin LAr presampler covering
\(|\eta| < 1.8\) to correct for energy loss in material upstream of the calorimeters. Hadronic
calorimetry is provided by the steel/scintillator-tile calorimeter, segmented into three barrel
structures within \(|\eta| < 1.7\), and two copper/LAr hadronic endcap calorimeters. The solid angle
coverage is completed with forward copper/LAr and tungsten/LAr calorimeter modules optimised for
electromagnetic and hadronic energy measurements, respectively.

The muon spectrometer (MS) comprises separate trigger and high-precision tracking chambers measuring
the deflection of muons in a magnetic field generated by the superconducting air-core toroidal
magnets. The field integral of the toroids ranges between \num{2.0} and \qty{6.0}{\tesla\metre}
across most of the detector. Three layers of precision chambers, each consisting of layers of
monitored drift tubes, cover the region \(|\eta| < 2.7\), complemented by cathode-strip chambers in
the forward region, where the background is highest. The muon trigger system covers the range
\(|\eta| < 2.4\) with resistive-plate chambers in the barrel, and thin-gap chambers in the endcap
regions.

The luminosity is measured mainly by the LUCID--2~\cite{LUCID2} detector that records Cherenkov light
produced in the quartz windows of photomultipliers located close to the beampipe.

Events were selected by the first-level trigger system implemented in custom hardware, followed by
selections made by algorithms implemented in software in the high-level trigger~\cite{TRIG-2016-01}.
The first-level trigger accepted events from the \qty{40}{\MHz} bunch crossings at a rate close to
\qty{100}{\kHz}, which the high-level trigger further reduced in order to record complete events to
disk at about \qty{1.25}{\kHz}.

A software suite~\cite{SOFT-2022-02} is used in data simulation, in the reconstruction and analysis
of real and simulated data, in detector operations, and in the trigger and data acquisition systems
of the experiment.

% End of text imported from the .//atlas-detector.tex input file

%
\section{Data and simulated event samples}
\label{sec:data}
The analysis uses $140~\text{fb}^{-1}$ of proton--proton ($pp$) collision data collected at $\sqrt{s}=13~\text{TeV}$ by the ATLAS experiment between 2015 and 2018, corresponding to the full LHC Run~2.
Only events recorded under stable-beam conditions with all detector subsystems operating normally are included, and standard data-quality requirements are applied to ensure reliable reconstruction of physics objects~\cite{DAPR-2018-01}.
The average number of simultaneous \(pp\) interactions per bunch crossing (pile-up) over the full dataset, \(\langle \muPU \rangle\), is approximately 34.

The search targets signal models in which a scalar mediator $S$ is produced via gluon--gluon fusion and subsequently decays into a quasi-conformal hidden sector,
following the framework described in Section~\ref{sec:intro}.
The mediator is taken to be either the SM Higgs boson with mass $m_S = 125$~GeV, or a heavier scalar particle with $m_S = 400$~GeV or $m_S = 750$~GeV.
The inclusion of heavier mediators allows the exploration of scenarios in which a larger total energy is injected into the hidden sector.
The dark-meson mass is set to the benchmark values $\mphi = 1.5$, $3$, and $5$~GeV, chosen as representative points within the range motivated in Section~\ref{sec:intro}.
Consistent with the expectation that $T$ is of the same order as $\mphi$, it is varied according to $T = \tfrac{2}{3}\mphi$, $T = \mphi$, and $T = \tfrac{3}{2}\mphi$.

Representative benchmark scenarios for \(\mAprime\) are defined within the region that kinematically allows decays to muons.
A dark-photon mass of $\mAprime = 0.5$~GeV, corresponding to branching fractions
$B(A'\rightarrow e^+e^-,\mu^+\mu^-,\pi^+\pi^-) = (0.4,\,0.4,\,0.2)$, is chosen as the nominal benchmark.
This choice yields a substantial muon fraction while retaining a non-negligible contribution from hadronic decays, and all combinations of $m_S$, $\mphi$, and $T$ are simulated for this benchmark.
To probe the dependence on the final-state composition, two alternative benchmarks are considered.
An enhanced-muon scenario with $\mAprime = 0.3$~GeV and
$B(A'\rightarrow e^+e^-,\mu^+\mu^-,\pi^+\pi^-) = (0.5,\,0.5,\,0.0)$
is studied for the case in which the mediator is the Higgs boson.
In addition, a reduced-muon scenario with $\mAprime = 0.7$~GeV and
$B(A'\rightarrow e^+e^-,\mu^+\mu^-,\pi^+\pi^-) = (0.15,\,0.15,\,0.7)$
is considered for a subset of model-parameter combinations.
A summary of the model parameters explored in this paper is shown in Table~\ref{tab_signal_models}.
Not all combinations of model parameters were generated, and the selection balances the exploration
of key model parameters and expectation of sensitivity to the chosen benchmarks.

\begin{table}[ht]
\caption{Benchmark signal model parameters considered. When multiple comma-separated values are given, all combinations across columns of those parameters were generated.}
\label{tab_signal_models}
\centering
\begin{tabular}{|c|c|c|c|}
\hline
$m_S$ (GeV) & $m_{\phi}$ (GeV) & $T$ (GeV) & \mAprime (GeV)\\
\hline
125 & 1.5 & 1.00, 1.50, 2.25 & 0.3, 0.5\\
125 & 3.0 & 2.0, 3.0, 4.5 & 0.5\\
125 & 5.0 & 3.33, 5.0, 7.5 & 0.3, 0.5\\
\hline
400 & 1.5 & 1.00, 1.50, 2.25 & 0.5, 0.7\\
400 & 3.0 & 2.0, 3.0, 4.5 & 0.5, 0.7\\
400 & 5.0 & 3.33, 5.0, 7.5 & 0.5, 0.7\\
\hline
700 & 1.5 & 1.00, 1.50, 2.25 & 0.5, 0.7\\
700 & 3.0 & 2.0, 3.0, 4.5 & 0.5, 0.7\\
700 & 5.0 & 3.33, 5.0, 7.5 & 0.5, 0.7\\
\hline
\end{tabular}
\end{table}

Signal Monte Carlo (MC) samples are generated at leading order (LO) using \PYTHIA[8.244.p3]~\cite{Mrenna:2016sih} with the A14 tune~\cite{ATL-PHYS-PUB-2014-021} and the \NNPDF[2.3lo]~\cite{Ball:2014uwa} parton distribution function (PDF) set.
Their inclusive gluon-fusion production cross-section is set to the Next-to-Next-Leading-Order (NNLO) QCD calculations including exact next-to-leading-order (NLO) top-mass effects~\cite{LHCHiggsCrossSectionWorkingGroup:2016ypw}.
The Higgs boson decay is handled by a dedicated Pythia plug-in based on the model described in Ref.~\cite{Knapen:2021eip}, which models the quasi-conformal shower in the dark sector.
The mediator branching ratio to the dark sector, $\brSUEP$, is set to 100\%.
Dark photons are then decayed with the branching ratios reported above.
The generated events are processed using a fast detector simulation based on a parameterization of the calorimeter response~\cite{SOFT-2010-01}.

For signal models in which the mediator is the SM Higgs boson, events are reweighted to reproduce the NLO prediction for the Higgs boson transverse momentum spectrum,
obtained from a sample generated with the \POWHEGBOX[v2] MC generator~\cite{Hamilton:2013fea,Hamilton:2015nsa,Alioli:2010xd,Nason:2004rx,Frixione:2007vw}.

A \zmumu MC sample is used to calibrate both the trigger efficiency and the track multiplicity correction described in Section~\ref{sec:reco}.
The \POWHEGBOX[v1] MC generator~\cite{Nason:2004rx,Frixione:2007vw,Alioli:2010xd,Alioli:2008gx} was used for the simulation of the
hard-scattering processes of $Z$-boson production at NLO in QCD, and for decaying it to the muon channel.
It was interfaced to \PYTHIA[8.186]~\cite{Sjostrand:2007gs} for the modeling of the parton shower, hadronization, and underlying event, with parameter values set according to the \AZNLO tune~\cite{STDM-2012-23}.
The \CT[10nlo] PDF set~\cite{Lai:2010vv} was used for the hard-scattering processes, whereas the \CTEQ[6L1] PDF set~\cite{Pumplin:2002vw} was used for the parton shower.
The effect of additional photons in the final state was simulated with \PHOTOSpp[3.52]~\cite{Golonka:2005pn,Davidson:2010ew}. The \EVTGEN[1.2.0] program~\cite{Lange:2001uf} was used to decay bottom and charm hadrons.

Together with the \zmumu sample, MC samples of \ttbar, \ttZ\, and diboson (\(VV\)) production are used to verify that only QCD multijet events are expected to contribute significantly to the search and validation regions defined in Section~\ref{sec_background}.
The production of \ttbar events was modeled using the
\POWHEGBOX[v2]~\cite{Frixione:2007nw,Nason:2004rx,Frixione:2007vw,Alioli:2010xd}
generator at NLO with the \NNPDF[3.0nlo]~\cite{Ball:2014uwa} PDF set
and the \hdamp parameter\footnote{The
\hdamp parameter is a resummation damping factor and one of the
parameters that controls the matching of \POWHEG matrix elements to
the parton shower and thus effectively regulates the
high-\pT radiation against which the \ttbar system recoils.} set
to 1.5\,\mtop~\cite{ATL-PHYS-PUB-2016-020}.  The events were interfaced
to \PYTHIA[8.230]~\cite{Sjostrand:2014zea} to model the parton shower,
hadronization, and underlying event, with parameters set according
to the A14 tune~\cite{ATL-PHYS-PUB-2014-021} and using the \NNPDF[2.3lo]
set of PDFs~\cite{Ball:2012cx}. The decays of bottom and charm hadrons
were performed by \EVTGEN[1.6.0]~\cite{Lange:2001uf}.

The production of \ttZ\ events was modeled using the
\MGNLO[2.3.3]~\cite{Alwall:2014hca} generator at NLO with the
\NNPDF[3.0nlo]~\cite{Ball:2014uwa} PDF.
The events were interfaced to \PYTHIA[8.210]~\cite{Sjostrand:2014zea}
using the A14 tune~\cite{ATL-PHYS-PUB-2014-021} and the
\NNPDF[2.3lo]~\cite{Ball:2014uwa} PDF set. The decays of bottom and charm
hadrons were simulated using the \EVTGEN[1.2.0] program~\cite{Lange:2001uf}.

Samples of diboson final states (\(VV\)) were simulated with the
\SHERPA[2.2.1] or 2.2.2~\cite{Bothmann:2019yzt} generator depending on the process,
including off-shell effects and Higgs boson contributions, where appropriate.
Fully leptonic final states and semileptonic final states, where one boson
decays leptonically and the other hadronically, were generated using
matrix elements at NLO accuracy in QCD for up to one additional parton
and at LO accuracy for up to three additional parton
emissions. Samples for the loop-induced processes \(gg \to VV\) were
generated using LO-accurate matrix elements for up to one
additional parton emission for both the cases of fully leptonic and
semileptonic final states. The matrix element calculations were matched
and merged with the \SHERPA parton shower based on Catani--Seymour
dipole factorization~\cite{Gleisberg:2008fv,Schumann:2007mg} using the MEPS@NLO
prescription~\cite{Hoeche:2011fd,Hoeche:2012yf,Catani:2001cc,Hoeche:2009rj}.
The virtual QCD corrections were provided by the
\OPENLOOPS library~\cite{Buccioni:2019sur,Cascioli:2011va,Denner:2016kdg}. The
\NNPDF[3.0nnlo] set of PDFs was used~\cite{Ball:2014uwa}, along with the
dedicated set of tuned parton-shower parameters developed by the
\SHERPA authors.

All background MC samples are processed using a full \GEANT-based detector simulation~\cite{SOFT-2010-01}.
For all MC samples, pile-up was modeled by overlaying the simulated hard-scattering event with
inelastic \(pp\) events generated with \PYTHIA[8.186]~\cite{Sjostrand:2007gs}
using the \NNPDF[2.3lo] set of PDF~\cite{Ball:2012cx} and the
A3 set of tuned parameters~\cite{ATL-PHYS-PUB-2016-017}.
All MC samples are reweighted such that the distribution of the number of pile-up vertices reproduces that observed in data.

\section{Event reconstruction}
\label{sec:reco}
Events are selected using the multi-muon trigger with the lowest \(\pt\) threshold that collected the full integrated luminosity in each data-taking period.
This corresponds to a trigger requiring at least three muons with \(\pt > 6\)~GeV in 2015 and 2016, and at least four muons with \(\pt > 4\)~GeV in 2017 and 2018.
Simulated signal samples are corrected to reproduce the trigger efficiency expected in data.
A scale factor is applied to the simulated signal events, defined as the ratio of the trigger efficiency expected in data to that in simulation.
Following Ref.~\cite{TRIG-2018-01}, the expected multi-muon trigger efficiency is estimated from the corresponding single-muon trigger efficiencies
with identical transverse momentum thresholds, assuming that the efficiencies for individual muons are uncorrelated.

\begin{figure}
\centering
\subfloat[]{
\includegraphics[width=0.5\textwidth]{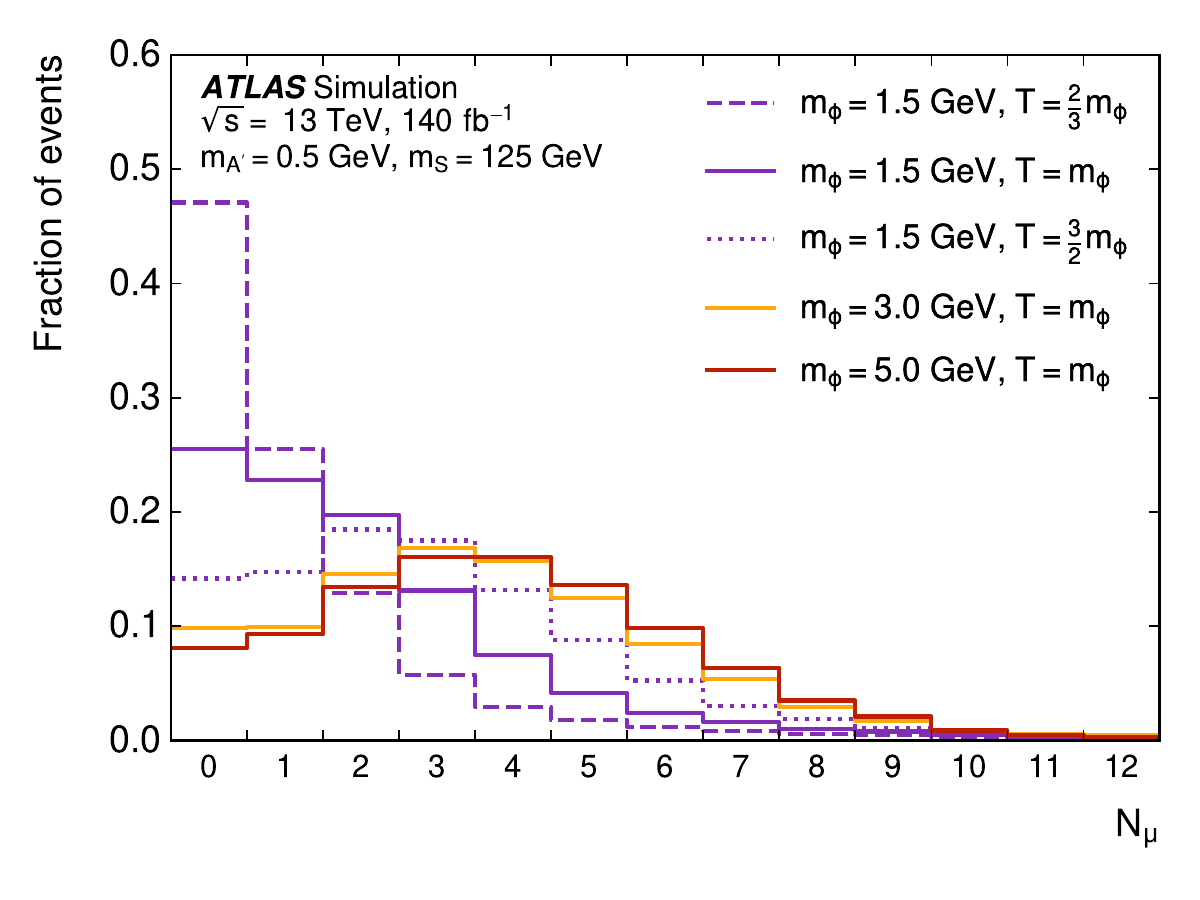}
\vspace{-2mm}
\label{fig_signal_kinematic_a}
\vspace{-2mm}
}
\subfloat[]{
\includegraphics[width=0.5\textwidth]{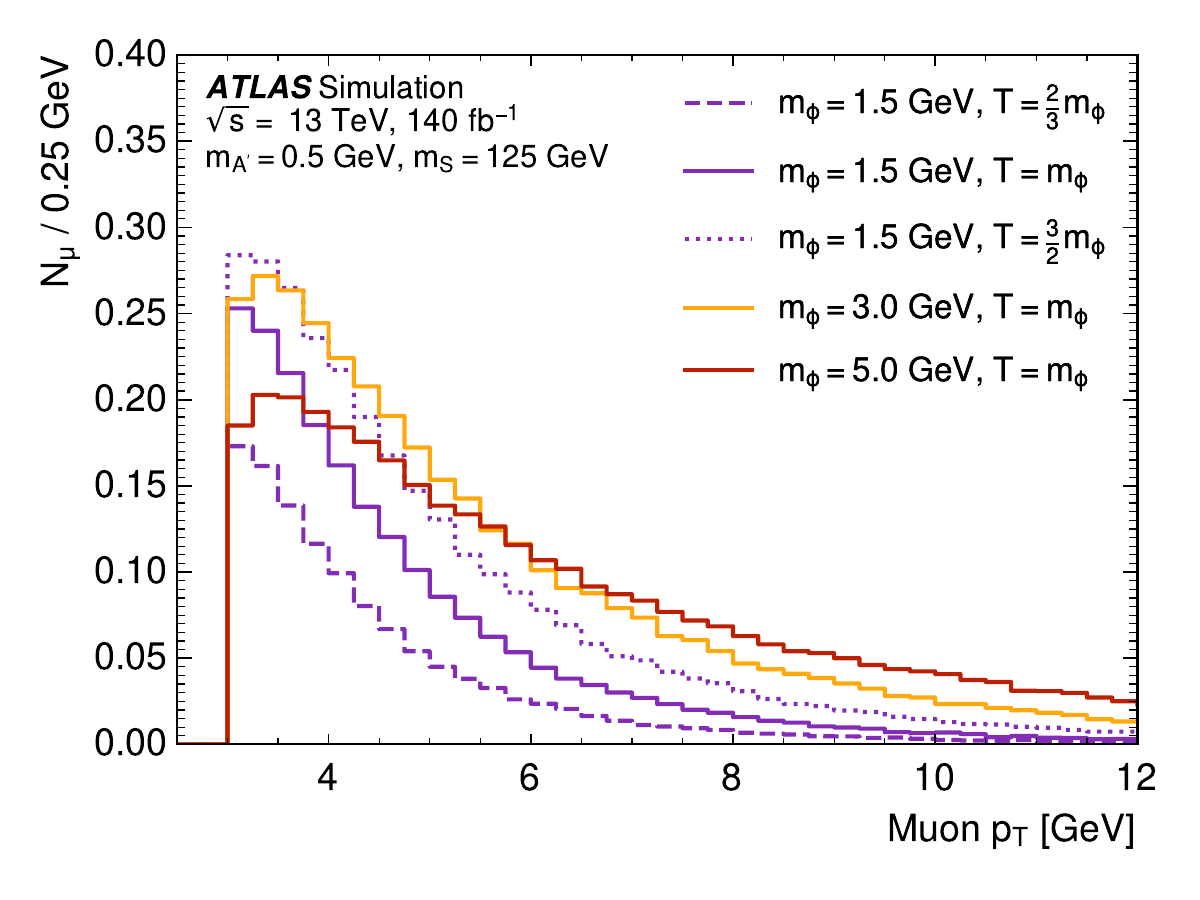}
\vspace{-2mm}
\label{fig_signal_kinematic_b}
\vspace{-2mm}
}
\caption{
Expected reconstructed \protect\subref{fig_signal_kinematic_a} muon multiplicity and \protect\subref{fig_signal_kinematic_b} muon \(\pt\) spectrum,
illustrating the dependence of these observables on the temperature \(T\) and dark-meson mass \(\mphi\) for a Higgs boson mediator and $\mAprime=0.5$~GeV.
The muon selection requirements of the analysis are applied, but no event selection or trigger requirements.
}
\label{fig_signal_kinematic}
\end{figure}

The single-muon efficiencies are measured in data and in the \zmumu MC sample for muons passing the offline selection requirements, using a tag-and-probe method similar to that described in Ref.~\cite{TRIG-2018-01}.
In addition to the binning in $\eta$, $\phi$, and data-taking period used in Ref.~\cite{TRIG-2018-01}, an additional binning in $\pt$ is introduced to better model the reduced efficiency in the trigger turn-on region.
The $\pt$ bin edges are defined as $[4,6,8,10,\infty)$~GeV.
At least three (four) muons reconstructed by the High Level Trigger are required to be within $\Delta R = 0.07$ of an offline muon in 2015 and 2016 (2017 and 2018).
The trigger efficiency ranges from approximately 1\% for all signal models with $\mphi=1.5$~GeV and $T=1$~GeV to nearly 100\% for the model with \(M=750\)~GeV, \(\mAprime=0.5\)~GeV, and $\mphi=T=5$~GeV.
The corresponding scale factors range between 0.85 and 1.
The variations in trigger efficiency and scale factor are driven by differences in the muon kinematics, in particular the \(\pt\) spectrum and multiplicity, between signal models.
Figure~\ref{fig_signal_kinematic} illustrates this dependence for muons passing the selection requirements detailed below, before any event selection is applied.

Events are required to have at least one reconstructed vertex~\cite{IDTR-2021-01} with at least two associated tracks
with \pT\ larger than 500~\MeV\ that are consistent with originating from the beam collision region.
The vertex with the highest sum of squared transverse momenta of associated tracks is considered as the primary vertex (PV) of the event.
The transverse and longitudinal impact parameters of all tracks, denoted by \(d_0\) and \(z_0\), respectively, are calculated relative to the beamline.
For the longitudinal impact parameter, the \(z\)-position of the primary vertex is used as the reference point.

Muons are reconstructed by combining tracks in the ID with tracks in the MS in the range $|\eta| < 2.5$.
They are required to have $\pt > 3~\GeV$ and to satisfy the cut-based \textit{LowPt} working point defined in Ref.~\cite{MUON-2018-03}.
The combined tracks must also meet the requirements $|d_0| < 2~\text{mm}$ and $|\Delta z_0 \sin\theta| < 3~\text{mm}$,
ensuring muons originate close to the PV and suppressing backgrounds with non-prompt decays from heavy-flavor hadrons.

ID tracks are used to characterize the charged-particle activity in the event.
These tracks include those associated with reconstructed muons as well as other charged particles.
They must satisfy $\pt > 0.5~\GeV$, $|\eta| < 2.5$, $|d_0| < 0.5~\text{mm}$, and $|\Delta z_0 \sin\theta| < 0.5~\text{mm}$,
ensuring they are within the ID acceptance, reconstructed by the track reconstruction algorithms, and compatible
with originating at the PV.

\begin{figure}
\centering
\includegraphics[width=0.5\linewidth]{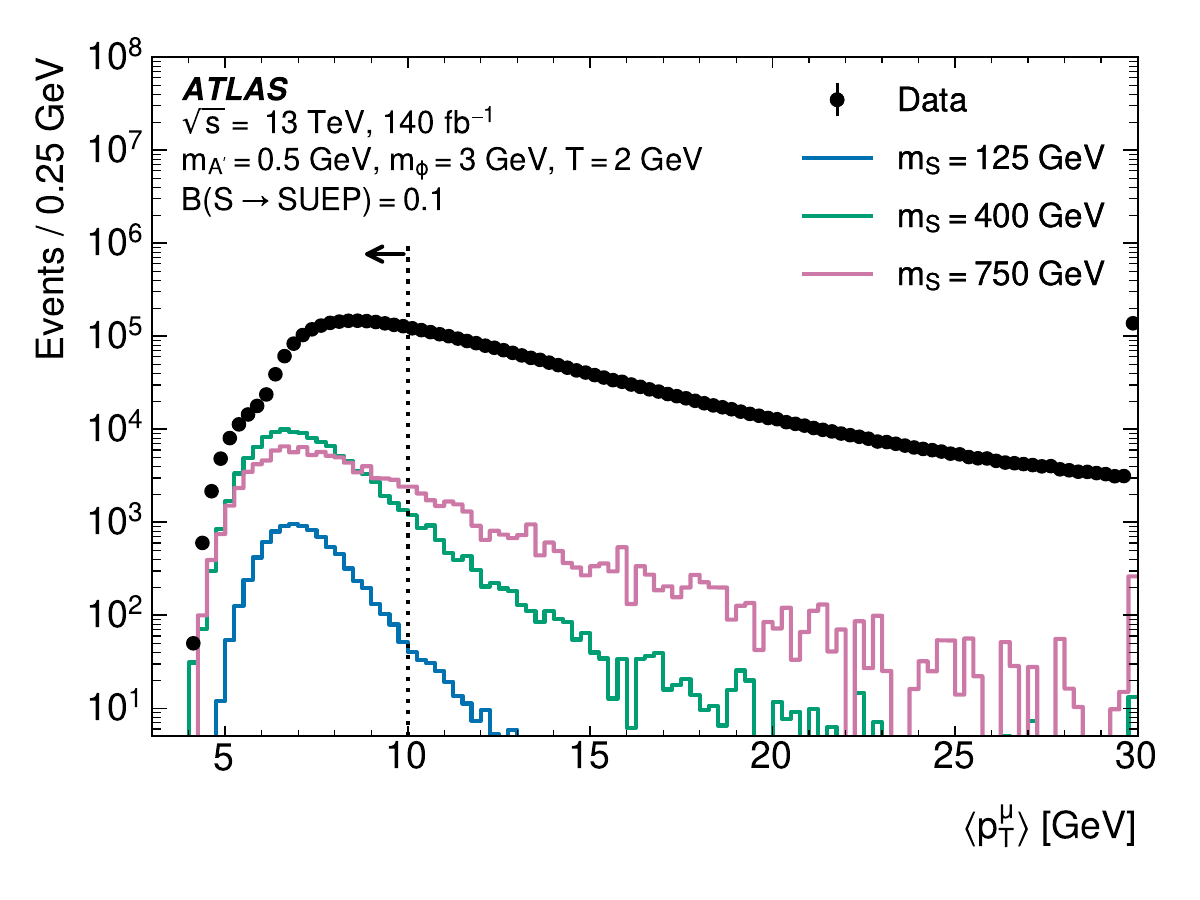}
\caption{
Average \(\pt\) of muons passing the selection requirements for events satisfying the trigger requirements.
Data are compared with a range of signal models to illustrate the dependence on the scalar mediator mass.
The dotted line and arrow indicate the region selected by the analysis.
Overflow entries are included in the last bin.
}
\label{fig_sig_bkg_kinematic}
\end{figure}

Events are required to have an average muon transverse momentum \(\avmuonpt < 10\)~GeV, computed using all muons that pass the selection requirements.
The distribution of \(\avmuonpt\) for data and selected signal models is shown in Figure~\ref{fig_sig_bkg_kinematic}.
This requirement strongly suppresses backgrounds in which muons originate from the decays of heavy particles such as vector bosons or top quarks.
According to simulation, after applying this requirement, the largest background contribution other than QCD multijet production arises from \ttbar\ production and amounts to approximately 0.5\% of events entering the search region,
while contributions from \zmumu, \ttZ, and \(VV\) processes are found to be at the level of \(10^{-4}\) or smaller.
Other processes involving heavy particles, such as \ttW, \ttH, and triboson production, are expected to contribute at a similar level.
As a result, the background is dominated by QCD multijet events, where muons are predominantly produced in decays of heavy-flavor hadrons, as well as kaons and pions.
As described in Section~\ref{sec_background}, the background contribution is estimated using a likelihood-based ABCD method based on the muon sphericity \(\sph\) and the track multiplicity \(\ntracks\), defined below.

\subsection{Sphericity}

Sphericity is an event-shape observable used to characterize the geometrical properties of particle production in high-energy collisions~\cite{Hanson:1975fe}.
It is typically computed using all reconstructed particles. In this analysis, it is instead constructed using only reconstructed muons.

\begin{figure}
\centering
\subfloat[]{
\includegraphics[width=0.5\linewidth]{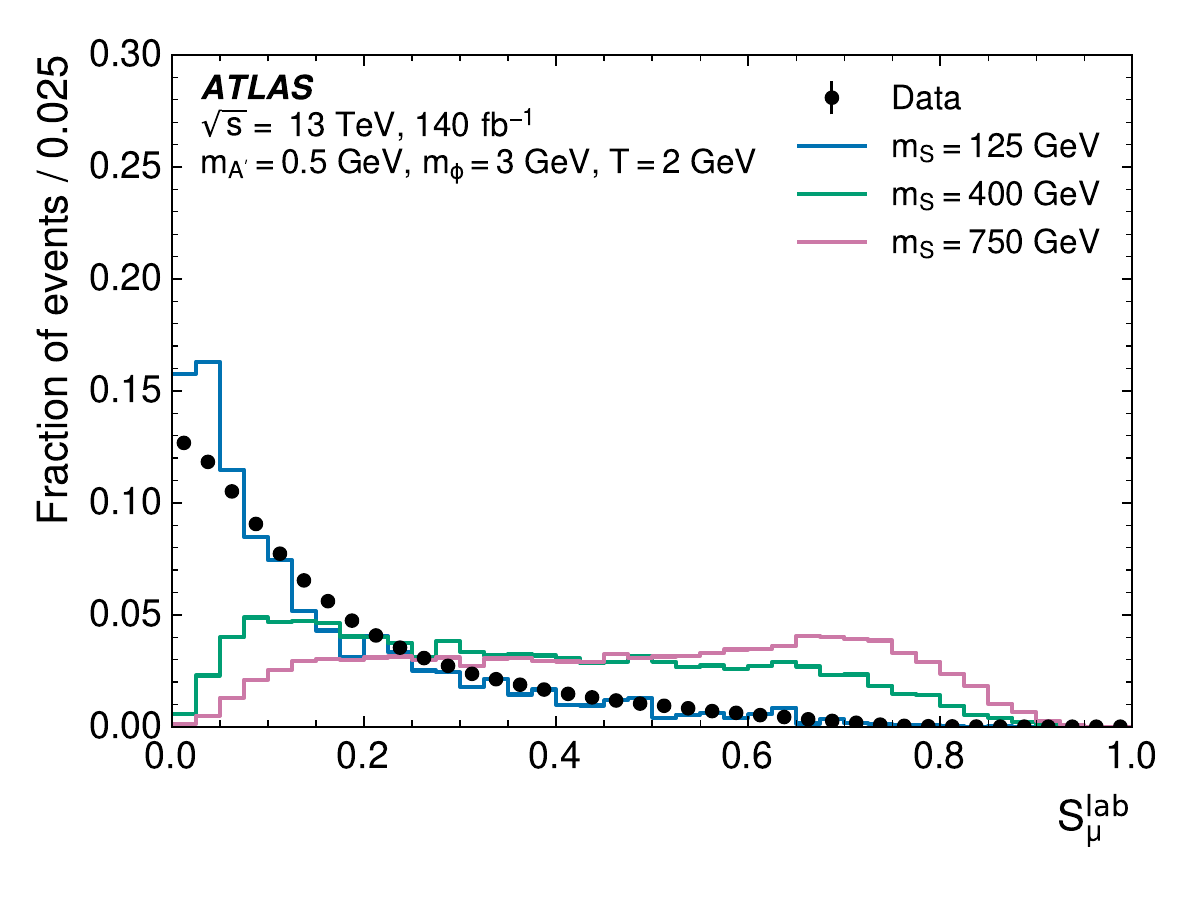}
\vspace{-2mm}
\label{fig_sphericity_a}
\vspace{-2mm}
}
\subfloat[]{
\includegraphics[width=0.5\linewidth]{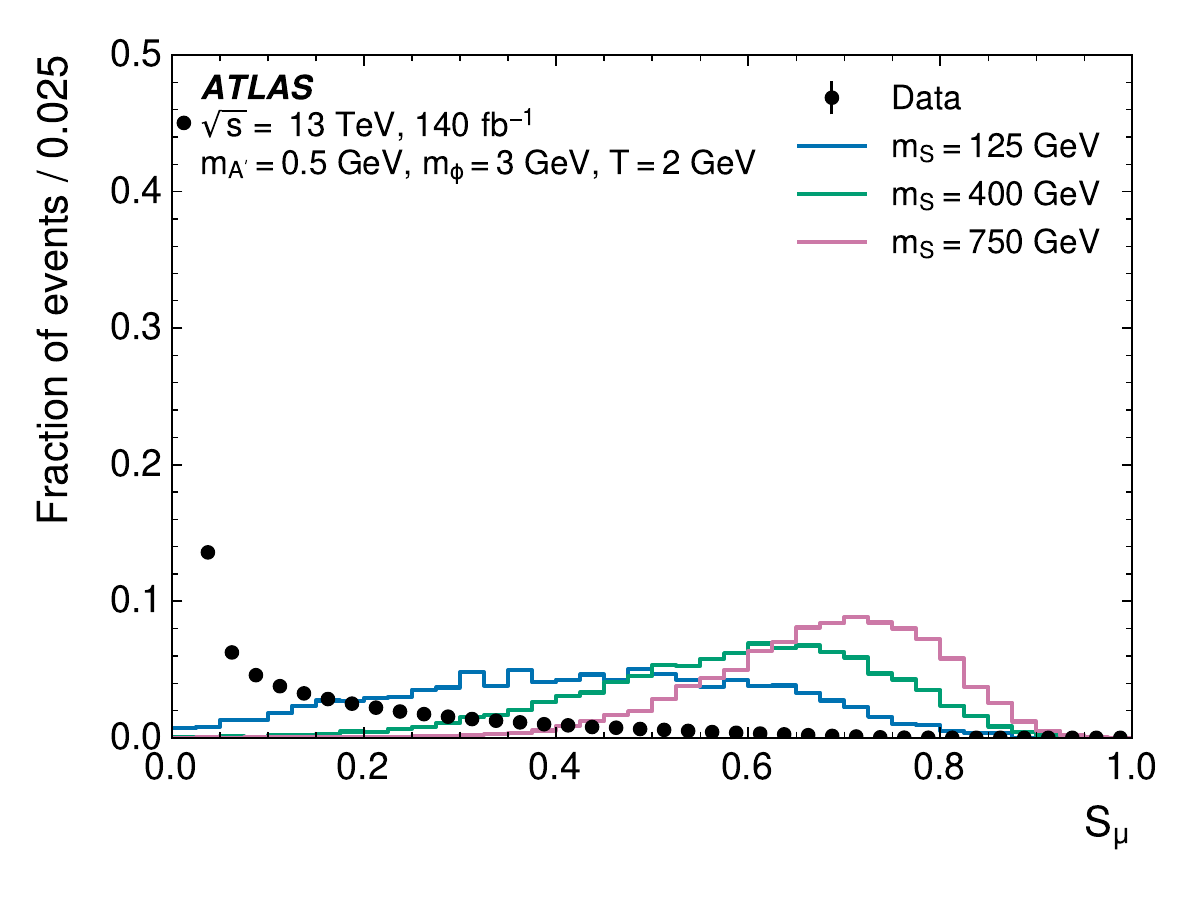}
\vspace{-2mm}
\label{fig_sphericity_b}
\vspace{-2mm}
}
\caption{
Sphericity of muons \protect\subref{fig_sphericity_a} computed in the laboratory frame $\sph^\textnormal{lab}$
and \protect\subref{fig_sphericity_b} computed after boosting to the muon-system rest frame $\sph$  for selected signal models and data.
The distributions are shown for events passing the trigger selection requirement and with \(\avmuonpt < 10\)~GeV.
}
\label{fig_sphericity}
\end{figure}

In signal events, the muons are approximately isotropically distributed in the rest frame of the mediator $S$, which is generally boosted in the laboratory frame,
particularly for models with low $m_S$, $\mphi$, or $T$ that rely on this boost to satisfy the trigger requirements.
To reduce the impact of this boost, all muon four-momenta are first summed to define the total four-momentum of the system.
A Lorentz transformation is performed to express the muon four-momenta in the rest frame of the system.
In this frame, the spatial momentum components are used to build the quadratic momentum tensor,
\begin{equation}
S^{\alpha\beta} = \frac{\sum_i p_i^\alpha p_i^\beta}{\sum_i |\vec{p}_i|^2},
\end{equation}
where the sum runs over all selected muons and $\alpha,\beta$ denote spatial indices. The tensor is diagonalized to obtain its eigenvalues $\lambda_1 \ge \lambda_2 \ge \lambda_3$, which satisfy $\lambda_1+\lambda_2+\lambda_3=1$. The sphericity is then defined as
\begin{equation}
\sph = \frac{3}{2}(\lambda_2+\lambda_3).
\end{equation}
This observable takes values close to zero for very collimated configurations and approaches unity for isotropic configurations.
As shown in Figure~\ref{fig_sphericity}, for signal models the value of $\sph$ is larger than that obtained when computed in the laboratory frame, while in data the $\sph$ distribution is shifted towards lower values compared to the laboratory-frame definition.

\subsection{Track multiplicity}
\label{sec_ntracks}

\begin{figure}
\centering
\subfloat[]{
\includegraphics[width=0.5\linewidth]{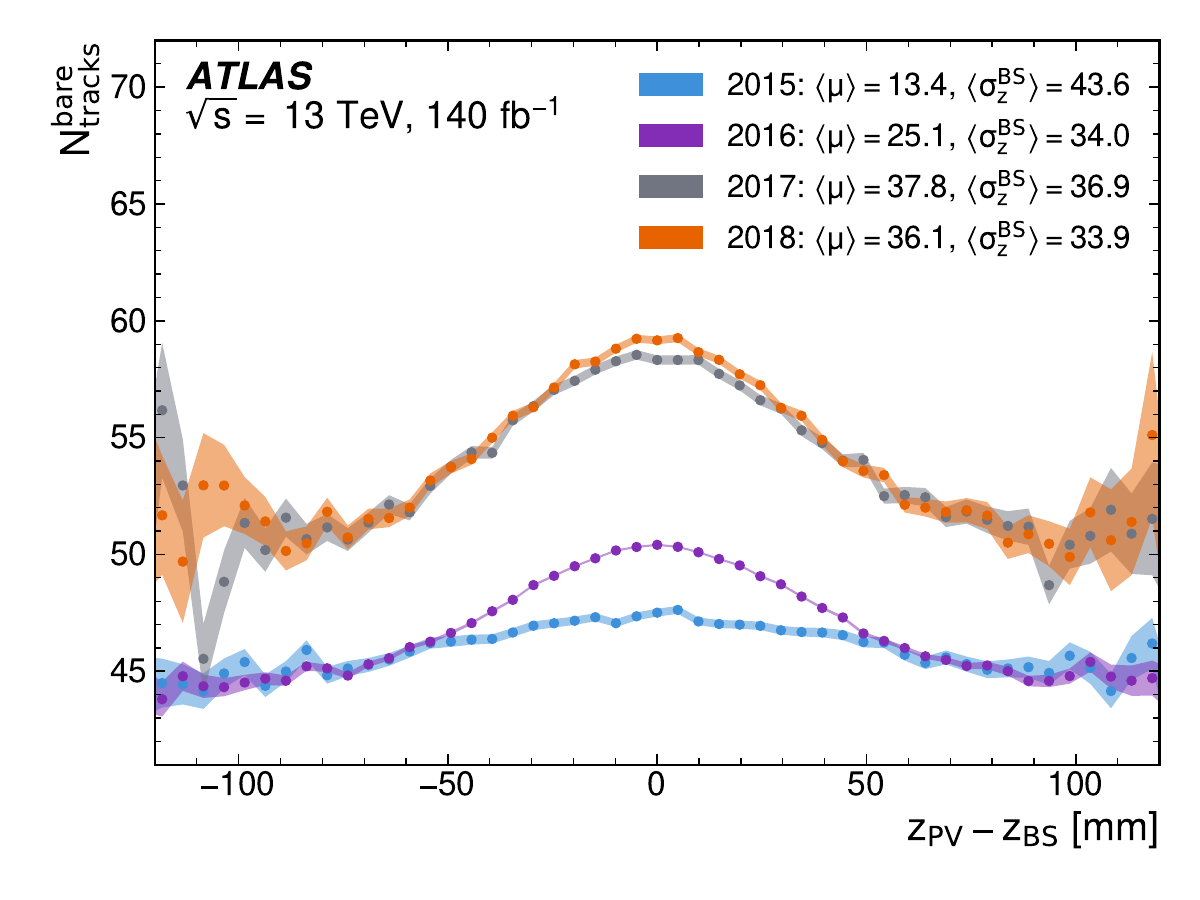}
\vspace{-2mm}
\label{fig_beamspot}
\vspace{-2mm}
}
\subfloat[]{
\includegraphics[width=0.5\linewidth]{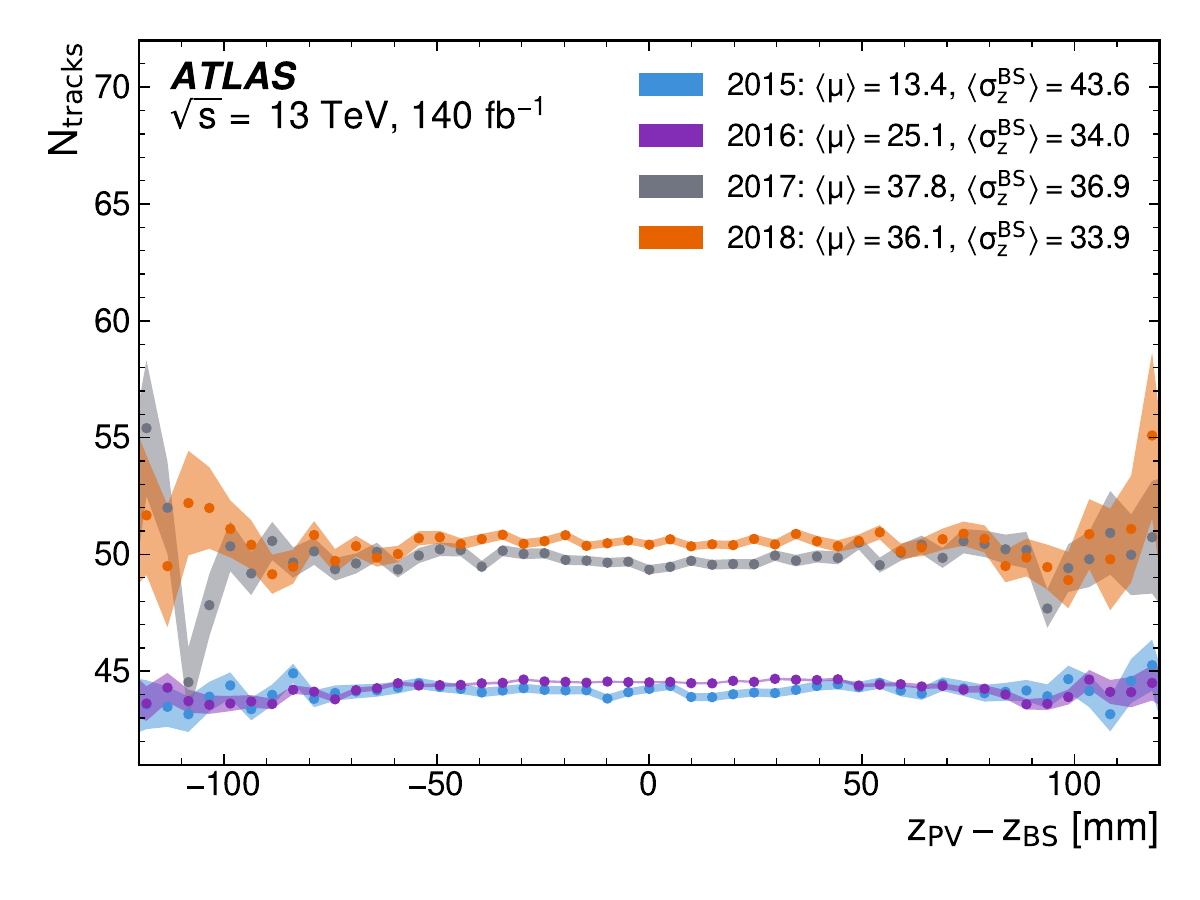}
\vspace{-2mm}
\label{fig_beamspot_corr}
\vspace{-2mm}
}
\caption{
\protect\subref{fig_beamspot} Bare and \protect\subref{fig_beamspot_corr} corrected track multiplicities
as a function of the longitudinal distance between the PV and the mean position of the luminous region,
shown for different years of data-taking. The error bands indicate the statistical uncertainty.
}
\end{figure}

The number of reconstructed tracks is a powerful observable for discriminating signal from background.
A large multiplicity of low-momentum muons,
electrons,\footnote{Electrons, which cannot be reliably reconstructed or identified in this analysis due to their low \(\pt\), as well as muons that are below the minimum \(\pt\) specified in Section~\ref{sec:reco}, may still be reconstructed as charged-particle tracks.}
and/or charged hadrons (depending on the specific $\mAprime$)
is expected in the case of signal events, far exceeding what is typical in background processes.

The tight requirement on the longitudinal track impact parameter ($|\Delta z_0 \sin\theta| < 0.5~\text{mm}$) helps
to isolate tracks originating from the hard-scattering interaction from those originating from pile-up interactions.
However, pile-up interactions can still contribute when their vertices are sufficiently close to the PV.
The size of this contribution depends on the local density of pile-up vertices near the PV and thus varies with data-taking conditions.
This behavior is illustrated in Figure~\ref{fig_beamspot}, which shows that the increase in the bare (uncorrected) track multiplicity \(\ntracksbare\) for
PVs near the longitudinal center of the luminous region (or beam spot) \(\zbs\) depends on the data-taking period, reflecting differences in the average number of pile-up interactions per bunch crossing
\(\muPU\) and in the longitudinal width of the beam spot \(\sigmabs\).
Additionally, significant uncertainties in the modeling of inelastic \(pp\) interactions affect the multiplicity and kinematic properties of charged particles produced in pile-up vertices.

To reduce the sensitivity of the analysis to these effects, a data-driven estimate of the average number of pile-up tracks passing the track selection is employed.
Such an estimate is expected to depend on beam conditions at the time the event was recorded and denoted \(\ntrackspu\), and it is subtracted from \(\ntracksbare\)
to define the corrected track multiplicity:
\begin{equation}\label{eq:ntracks}
\ntracks = \ntracksbare - \ntrackspu((\zpv-\zbs)/\sigmabs, \muPU).
\end{equation}
where $\ntrackspu$ is defined in Eq.~\ref{eq:ntrackspu} and \(\zpv\) is the longitudinal position of the primary vertex of the event.
The beam spot parameters \(\zbs\) and \(\sigmabs\) are determined from fits to the spatial distribution of
reconstructed primary vertices, accumulated over many events within a given time period of about 60~s~\cite{PERF-2015-01},
and \(\muPU\) is measured by the LUCID-2 detector~\cite{DAPR-2021-01}.

\begin{figure}
\centering
\includegraphics[width=0.5\linewidth]{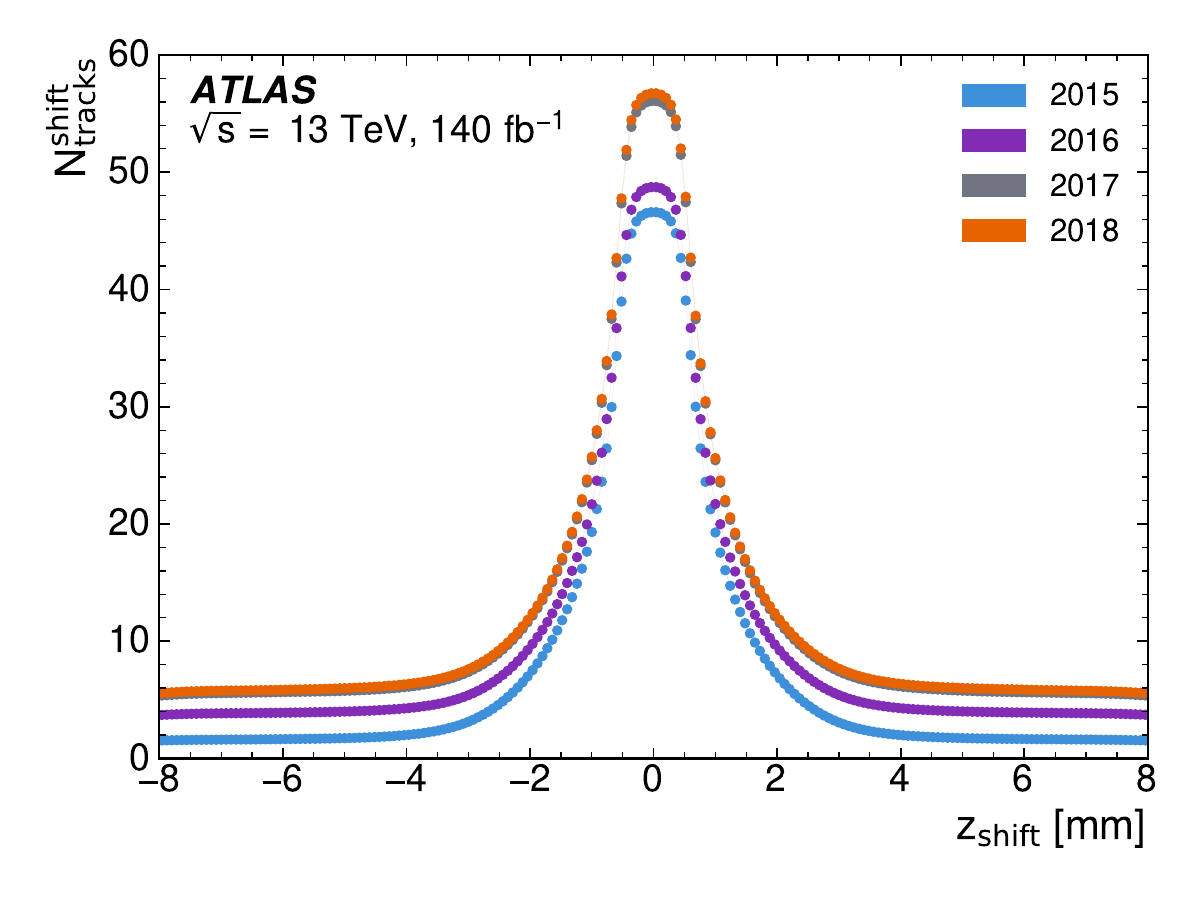}
\caption{Multiplicity of tracks $N_{\mathrm{tracks}}^{\mathrm{shift}}$ when taken as a function of the track longitudinal impact parameters with respect to the PV (\(z_{\mathrm{shift}}\)).}
\label{fig_scan}
\end{figure}

The contribution \(\ntrackspu\) is estimated using tracks whose longitudinal impact parameter is consistent with originating away from the PV,
where only pile-up contributions are expected.
Figure~\ref{fig_scan} shows the track multiplicity as a function of the longitudinal displacement relative to the PV, denoted \(z_{\mathrm{shift}}\).
The contribution from tracks originating from the PV becomes negligible for \(|z_{\mathrm{shift}}| \gtrsim 4\)~mm, as indicated by the approximately flat distribution in these ranges.
A mild decrease toward larger values of \(|z_{\mathrm{shift}}|\) is observed and expected, consistent with the reduced pile-up vertex density away from \(\zbs\).
The measurement of \(\ntrackspu\) is therefore performed by counting how many tracks are within a window of $1$~mm around \(z_{\mathrm{shift}} = 6.5\)~mm,
using events that satisfy the trigger requirements.

The resulting track multiplicity is dependent on $\muPU$ and on the location relative to the beamspot center,
normalized by the beamspot width (\((z-\zbs)/\sigmabs\)).
The expected \(\ntrackspu\) at any given longitudinal position $z$ can then be derived assuming a Gaussian dependence on \((z-\zbs)/\sigmabs\) and a linear dependence on \(\muPU\):
\begin{equation}\label{eq:ntrackspu}
\ntrackspu((z-\zbs)/\sigmabs, \muPU)
= \muPU\, A \exp\!\left[-\tfrac{1}{2}
\left(\frac{(z-\zbs)/\sigmabs}{B}\right)^2\right] + N_0 ,
\end{equation}
where \(A\), \(B\), and \(N_0\) are determined from a \(\chi^2\) fit to the measured track multiplicities.
The parameter \(N_0\) accounts for additional contributions that are largely independent of \(\muPU\) and \((z-\zbs)/\sigmabs\).
For consistency, the procedure is performed independently in data and simulation, using a \(\zmumu\) sample for the latter.
The fitted parameters for both data and simulation are listed in Table~\ref{table_ntracksPU}.

As shown in Figure~\ref{fig_beamspot_corr}, the corrected observable \(\ntracks\) exhibits no significant dependence on the PV position within the beam spot or on variations in instantaneous luminosity.
A residual difference between the 2015--2016 and 2017--2018 data-taking periods is observed and expected due to the different trigger selections employed during these periods,
leading to a different average track multiplicity from the PV.

\begin{table}[ht]
\caption{Fitted parameters of $\ntrackspu$ for data and MC with $\pm1\sigma$ fit uncertainties.}
\label{table_ntracksPU}
\centering
\begin{tabular}{llll}
\hline
& $A$ & $B$ & $N_0$  \\
\hline
Data & $0.2240 \pm 0.0006$ & $0.9589 \pm 0.0028$ & $0.474 \pm 0.005$ \\
MC & $0.2072 \pm 0.0006$ & $0.9182 \pm 0.0027$ & $0.440 \pm 0.005$ \\
\hline
\end{tabular}
\end{table}

\section{Event selection and background estimation}
\label{sec_background}

After the preselection requirement \(\avmuonpt < 10\)~GeV, the dominant background arises from QCD multijet events in which heavy-flavor hadrons decay into muons.
While the signal is assumed to be prompt, heavy-flavor hadrons can have sizable lifetimes, resulting in displaced muon signatures.
The search region therefore requires the transverse impact-parameter significance of the highest-\(\pt\) muon, \(\dsig = \left| \frac{d_0(\mu_1)}{\sigma_{d_0}(\mu_1)} \right| < 3\), where \(\mu_1\) denotes the highest-\(\pt\) muon.
In addition, a requirement on the muon multiplicity, \(\nmu \geq 5\), further suppresses the background.
The selection efficiency of the requirements on \(\avmuonpt\), \(\dsig\), and \(\nmu\) is close to unity for all signal models.

\begin{figure}
\centering
\subfloat[]{
\includegraphics[width=0.5\textwidth]{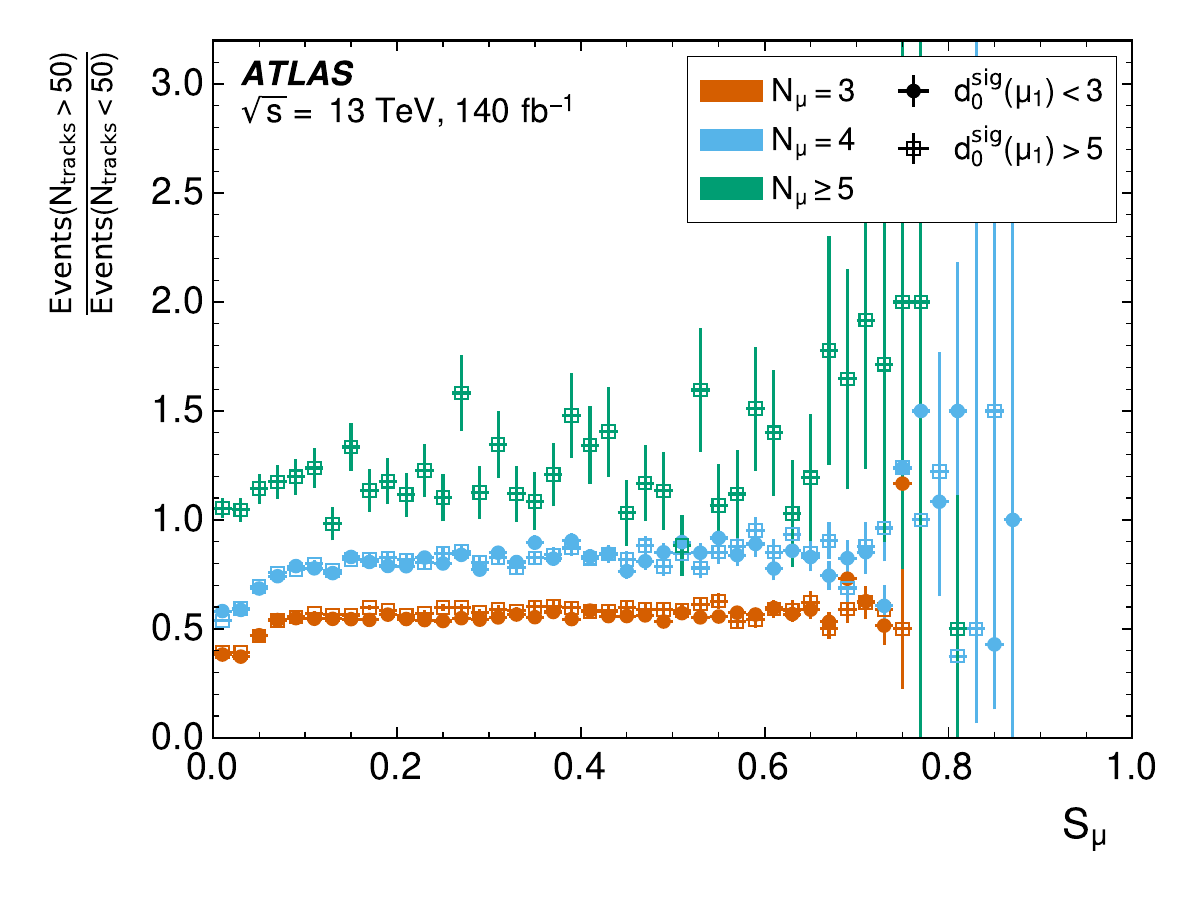}
\vspace{-2mm}
\label{fig_ratio_a}
\vspace{-2mm}
}
\subfloat[]{
\includegraphics[width=0.5\textwidth]{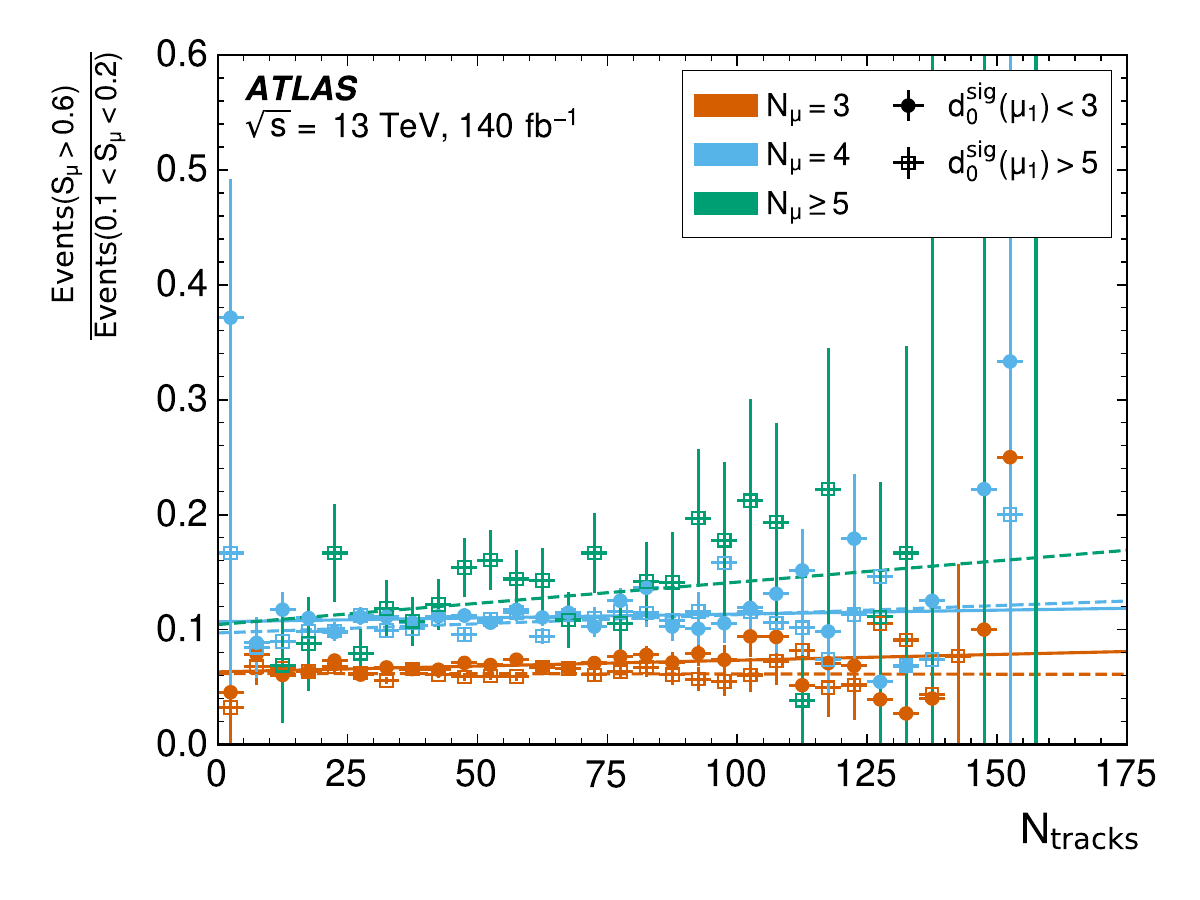}
\vspace{-2mm}
\label{fig_ratio_b}
\vspace{-2mm}
}
\caption{
\protect\subref{fig_ratio_a} Ratio of the number of data events with \(\ntracks > 50\) to those with \(\ntracks < 50\) as a
function of \(\sph\) and \protect\subref{fig_ratio_b} ratio of the number of data events with \(\sph > 0.6\) to those with \(0.1 < \sph < 0.2\)
as a function of \(\ntracks\), shown for each validation region.
The solid (dashed) lines in the right panel indicate linear fits to the ratio for validation
regions with \(\dsig < 3\) (\(\dsig > 5\)).
}
\label{fig_ratio}
\end{figure}

Background events with displaced leptonic decays are not expected to differ significantly in topology from those entering the search region.
Alternative requirements on \(\dsig\) and \(\nmu\) are therefore used to define regions in which the background estimation method is tested and validated.
The selections defining the validation and search regions are summarized in Table~\ref{table_selection_requirement}.

\begin{table}[ht]
\caption{Definition of the validation and search regions.}
\label{table_selection_requirement}
\centering
{\renewcommand{\arraystretch}{1.4}
\begin{tabular}{|c|c|c|}
\hline
\multicolumn{2}{|c|}{Requirements} & Region purpose \\
\hline
\multirow{2}{*}{$\nmu=3$}
& $\dsig > 5$ & validation \\
& $\dsig < 3$ & validation \\
\hline
\multirow{2}{*}{$\nmu=4$}
& $\dsig > 5$ & validation \\
& $\dsig < 3$ & validation \\
\hline
\multirow{2}{*}{$\nmu\geq 5$}
& $\dsig > 5$ & validation \\
& $\dsig < 3$ & search \\
\hline
\end{tabular}}
\end{table}

As illustrated in Figure~\ref{fig_ratio}, \(\ntracks\) and \(\sph\) are largely uncorrelated in the validation regions once the region $\sph < 0.1$ is excluded.
This low-\(\sph\) region is dominated by events in which muons are collimated into two narrow cones with well-aligned axes.
Such topologies are predominantly expected in events without additional jet activity away from this axis, resulting in a significantly
reduced \(\ntracks\) and a stronger correlation between \(\ntracks\) and \(\sph\).
A small trend in the ratio of the number of events satisfying \(\sph > 0.6\) to those with \(0.1 < \sph < 0.2\) is observed,
particularly pronounced in the validation region defined by \(\nmu \geq 5\) and \(\dsig > 5\). A dedicated systematic uncertainty
is discussed in Section~\ref{sec:sys} that takes this correlation into account.

\begin{figure}[t]
\centering
\subfloat[]{
\includegraphics[width=0.5\linewidth]{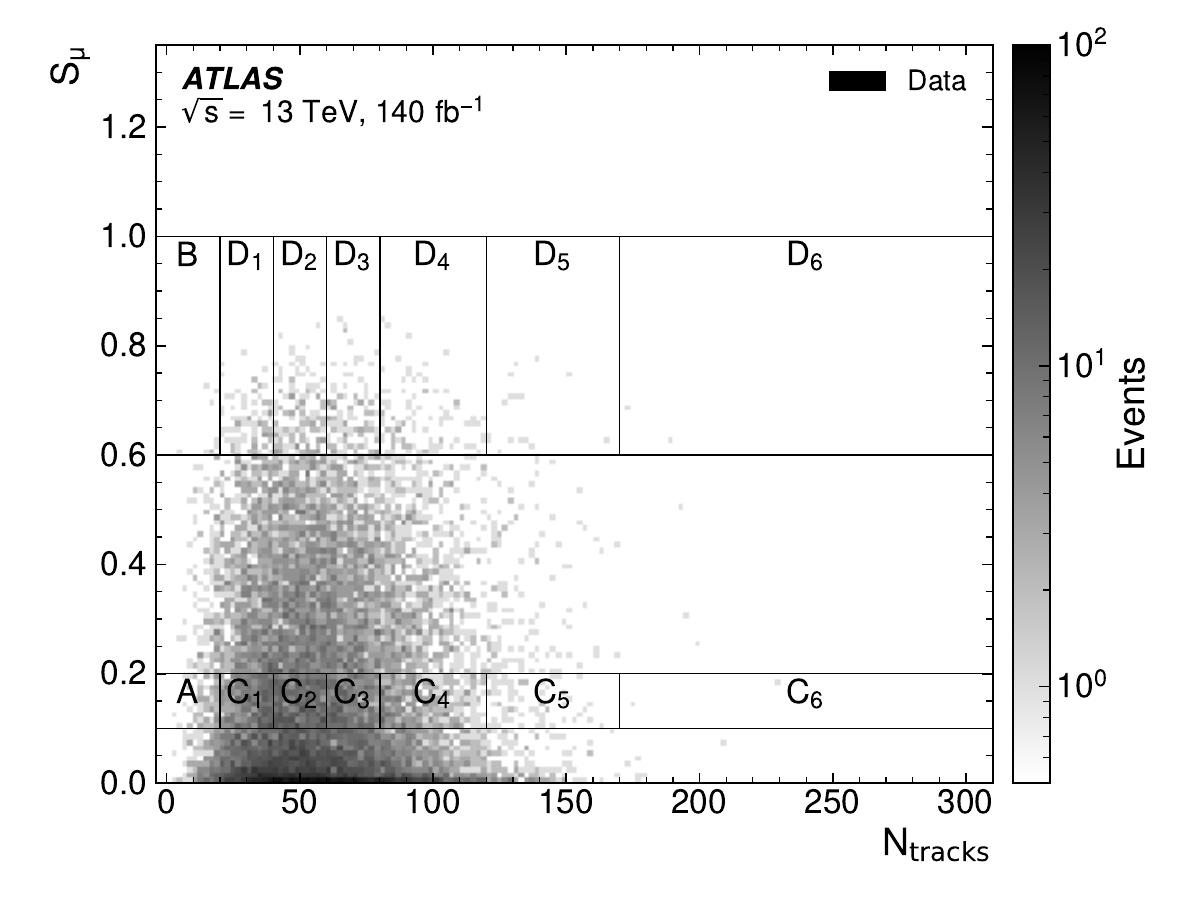}
\vspace{-2mm}
\label{figure_ABCD_method_a}
\vspace{-2mm}
}
\subfloat[]{
\includegraphics[width=0.5\linewidth]{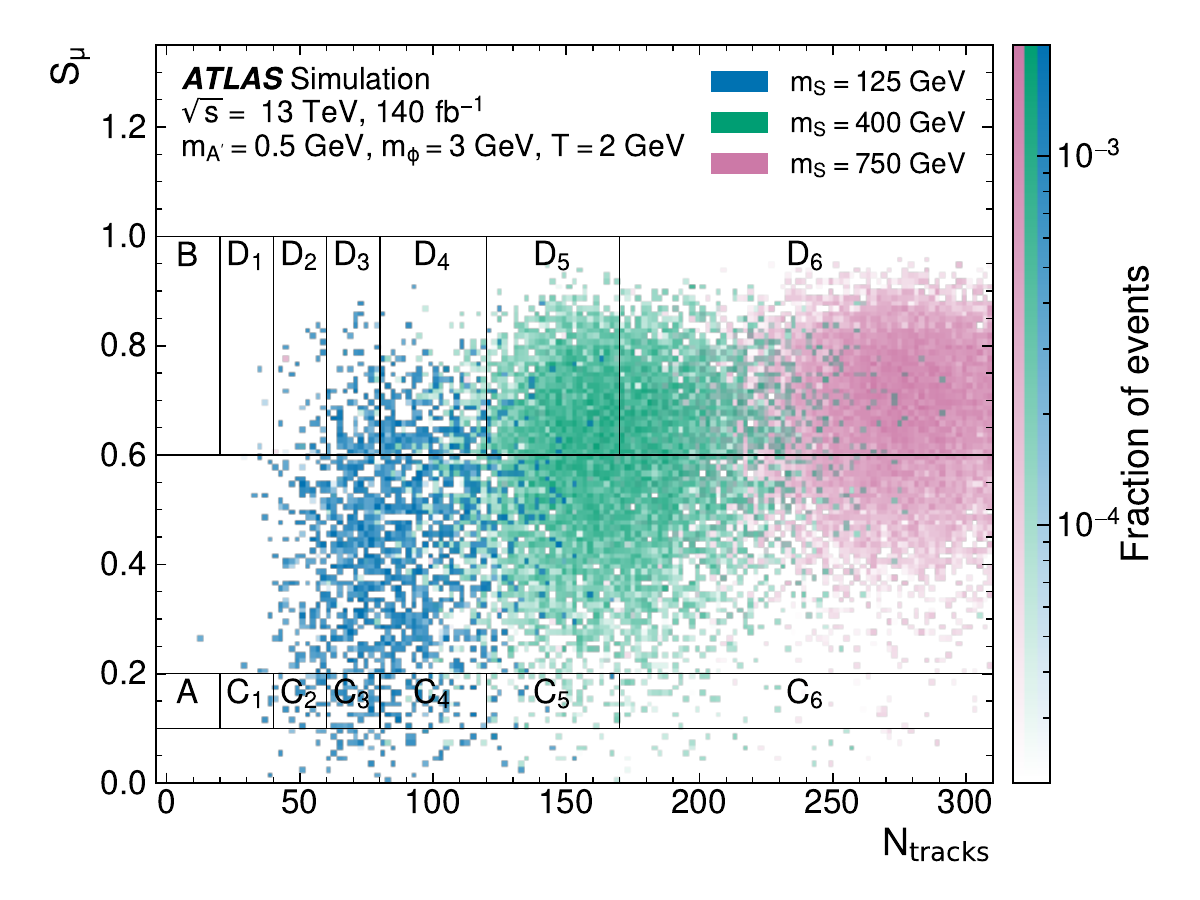}
\vspace{-2mm}
\label{figure_ABCD_method_b}
\vspace{-2mm}
}
\caption{Binning of the $\ntracks$--$\sph$ histograms and distributions in the search region of \protect\subref{figure_ABCD_method_a} data
and \protect\subref{figure_ABCD_method_b} selected signal models.
The signal entries are normalized to unity. The signal region corresponds to the bins \(D_1,\,\ldots,\,D_6\).}
\label{figure_ABCD_method}
\end{figure}

Based on the assumption that the correlation between \(\ntracks\) and \(\sph\) is similarly low in the search region,
the background is modeled using a multi-bin likelihood-based ABCD method.
The method allows a fully data-driven background prediction that does not rely on the simulated background samples.
Events are categorized in a two-dimensional histogram of \(\ntracks\) and \(\sph\), as shown in Figure~\ref{figure_ABCD_method}.
Along the \(\ntracks\) axis, the bin edges are defined at
$(-\infty,\,20,\,40,\,60,\,80,\,120,\,170,\,\infty)$,
with each bin inclusive of its lower boundary.
This binning is chosen to ensure good sensitivity across the full range of signal models considered, which show a wide variety of \(\ntracks\) distributions.
The \(\sph\) axis is partitioned into two disjoint intervals:
$0.1 < \sph < 0.2$ and \(\sph > 0.6\).

For increasing \(\ntracks\), the bins in the low-\(\sph\) interval are labelled
\(A,\,C_1,\,\ldots,\,C_6\),
while the corresponding bins in the high-\(\sph\) interval are labelled
\(B,\,D_1,\,\ldots,\,D_6\).
In the search region,
the background is expected to populate predominantly the control-region bins
\(A\), \(B\), and \(C_1,\,\ldots,\,C_6\),
whereas signal events are expected to contribute primarily to the signal region bins \(D_1,\,\ldots,\,D_6\).

The stringent requirement of \(\sph > 0.6\) defining the signal-region bins is chosen to enhance the sensitivity of the search relative to looser \(\sph\) selections.
Conversely, the upper bound \(\sph < 0.2\) for the control-region bins
\(A,\,C_1,\,\ldots,\,C_6\)
is chosen to suppress potential contamination from certain signal models,
allowing this region, together with the region $B$, to be unblinded during the development of the analysis strategy.
Since including the intermediate region \(0.2 < \sph < 0.6\) as additional signal-region bins does not improve the overall sensitivity,
this region is excluded from the analysis entirely.

The background model is constrained using maximum-likelihood fits to data, as further detailed in Section~\ref{sec:result}.
When testing signal hypotheses, the likelihood includes a signal component, such that the background normalization parameters are determined simultaneously with the scale of the signal contribution.
This procedure ensures a consistent determination of the background model both in background-only fits and in fits allowing for a potential signal contribution,
including cases where non-negligible signal contributions are present in control-region bins.

\section{Systematic uncertainties}
\label{sec:sys}
Experimental uncertainties on the simulated signal yields are evaluated to account for potential differences between data and simulated signal samples in the event reconstruction and selection.
The relative importance of the various uncertainty sources depends strongly on the signal model parameters.
For models with \(m_{\phi},\,T \lesssim 3\)~GeV, where the acceptance is largely constrained by the \pt thresholds of the online muon reconstruction, the dominant contribution is expected to arise from the trigger efficiency.
For models with \(m_{\phi},\,T \gtrsim 3\)~GeV, uncertainties associated with ID tracking are expected to become largest.
This behavior is reflected in the ranges quoted for the individual uncertainty sources below.

Two sources of systematic uncertainty affecting the trigger efficiency are evaluated, both arising from the tag--and--probe procedure used to determine the single-muon trigger efficiencies.
The first originates from the limited statistical precision of the measurement, with its impact on the signal yield in the search region ranging from about 0.1\% to 20\%, depending on the model parameters.
The second is derived by varying the selection requirements applied to the probe muon.
The set of variations follows Ref.~\cite{TRIG-2018-01}, except for the probe-muon \pt requirement, which is adapted to accommodate the additional \pt binning used.
The effect of this second source on the signal yield in the search region ranges from roughly 0.1\% to 4\%, depending on the model parameters.

Offline muon reconstruction is affected by uncertainties in the resolution of the combined muon track and in the reconstruction efficiency.
The impact of the muon track-resolution uncertainty is found to be negligible.
The uncertainty associated with the mismodeling of the muon reconstruction efficiency leads to variations in the signal yield in the search region ranging from approximately 0.5\% to 5\%, depending on the model parameters.

An uncertainty of 100\% is assigned to the estimated rate of tracks arising from combinatorial background in the ID, based on comparisons between data and simulation
of the reconstructed track multiplicity as a function of pile-up~\cite{IDTR-2021-01}.
Since the contribution of such tracks is small, its effect on $\ntracks$ ranges from approximately 1\% to 5\%, depending on the model parameters.
For models predicting very high $\ntracks$, this uncertainty is largest in relative terms; however, its impact on the search sensitivity is sub-dominant,
since all events remain within the highest $\ntracks$ bin under any variation.
Uncertainties on the ID track-resolution and reconstruction-efficiency modeling are found to be negligible for all signal models.

A pile-up modeling uncertainty is assigned to account for the difference between the predicted and measured inelastic cross-sections~\cite{STDM-2015-05}.
Its effect is evaluated to be below 3\% on the signal yield in the search region and below 1\% on $\ntracks$ for all signal models.

An uncertainty on $\ntrackspu$ is evaluated by repeating the measurement using either an alternative \(\zmumu\)-enriched event selection or alternative offsets of \(z_{\mathrm{shift}} = -6.5\)~mm and \(z_{\mathrm{shift}} = 7\)~mm.
The envelope of the variations is taken as the \(\pm 1\sigma\) uncertainty. The impact of this uncertainty on $\ntracks$ is found to be below 0.5\% for all signal models.

For signal models in which the mediator is the Higgs boson, acceptance uncertainties are evaluated by varying the renormalization and factorization scales, PDF eigenvectors and PDF sets, parton shower settings,
and by comparing the nominal NLO Higgs \(\pt\) spectrum prediction with an alternative NNLO prediction and its variations.
The combined uncertainty from these sources is evaluated to be 20\% and is applied to the expected signal yields in the \(D_i\) bins.

The uncertainty on the combined 2015--2018 integrated luminosity is 0.83\%~\cite{DAPR-2021-01}, derived using the LUCID-2 detector~\cite{LUCID2} for the primary luminosity measurements.

The statistical uncertainty of the regions used to estimate the background is propagated as a systematic uncertainty on the prediction.
Additionally, a systematic uncertainty of 5--10\% is assigned to the background yields in the \(D_i\) bins to account for the sensitivity of the prediction to the definition of the control regions.
This uncertainty is evaluated by varying both the lower and upper \(\sph\) boundaries of the \(A\) and \(C_i\) bins by \(\pm 0.01\), corresponding to small changes around the nominal region.

As illustrated in Figure~\ref{fig_ratio}, the ratio of the number of events satisfying \(\sph > 0.6\) to those with \(0.1 < \sph < 0.2\) increases slightly with increasing \(\ntracks\).
This trend is particularly pronounced in the validation region defined by \(\nmu \geq 5\) and \(\dsig > 5\).
To account for this residual correlation, the background model is allowed to deviate from the assumption of independence between \(\ntracks\) and \(\sph\), by an amount determined in the validation regions.
The ratio shown in Figure~\ref{fig_ratio} is fit with a linear function in each validation region, and the largest observed deviation from the expectation
of no correlation, corresponding to a zero slope, is incorporated as a systematic uncertainty in the background estimate.
Several possible sources of this positive correlation have been investigated. Those include slightly different composition in exclusive muon-multiplicity bins compared with the inclusive five or more muons required in the signal region,
a data-taking time dependence due to slightly different triggers active in 2015--2016 compared to 2017--2018, contributions from decays in flight of light charged hadrons (which were found to increase with the muon multiplicity), and
effects of correlated muon production through resonances as in \jpsimumu decays. None of these effects alone could fully explain the observed correlation,
although each contributed to some extent.

\section{Statistical analysis and results}
\label{sec:result}
The statistical interpretation is based on profile-likelihood-ratio test statistics.
The goodness of fit of the background-only model is evaluated using a likelihood-ratio test with respect to a model that perfectly describes the observed data,
while each signal hypothesis is tested using profile-likelihood ratios comparing the background-only and signal-plus-background hypotheses.
In all cases, \(p\)-values are computed using asymptotic approximations~\cite{Cowan:2010js}.

To account for the systematic uncertainty associated with the residual correlation between \(\ntracks\) and \(\sph\), described in Section~\ref{sec:sys}, the following relations are imposed on the expected background yields \(b_r\) in bin \(r\):

\begin{equation}
\begin{aligned}
b_A &= \mu_A, \\
b_B &= \mu_A \epsilon_{AB} \left(1 + \chi \langle \ntracks \rangle_B\right), \\
b_{C_i} &= \mu_A \epsilon_i, \\
b_{D_i} &= \mu_A \epsilon_{AB}\epsilon_i
\left(1 + \chi \langle \ntracks \rangle_{D_i}\right),
\qquad \forall\, i \in \{1,\ldots,6\},
\end{aligned}
\label{eq:abcd_like_lin}
\end{equation}

where \(\mu_A\), \(\epsilon_{AB}\), and \(\epsilon_i\) are normalization parameters,
\(\langle \ntracks \rangle_r\) denotes the average value of \(\ntracks\) in bin \(r\),
and \(\chi\) is a Gaussian-distributed nuisance parameter with mean zero (corresponding to the no-correlation case) and
standard deviation \(\sigma_{\chi}\), set as described below.

The parameter \(\chi\) allows deviations from the assumption that the ratios \(\frac{b_B}{b_A}\) and \(\frac{b_{D_i}}{b_{C_i}}\) are equal.
It captures a small linear dependence of these ratios on \(\ntracks\) that is characteristic of the background and not typically reproduced by potential signal contributions.
To determine \(\sigma_{\chi}\), linear fits to the ratio of the number of events with \(\sph > 0.6\) to those with \(0.1 < \sph < 0.2\) as a function of \(\ntracks\) are performed in the validation regions, as illustrated in Figure~\ref{fig_ratio}.
The value \(\sigma_{\chi} = 0.0036\) is taken as the ratio of the slope to the intercept obtained from the fit in the validation region defined by the requirements \(\nmu \geq 5\) and \(\dsig > 5\), where the slope is largest.

\begin{figure}
\centering
\subfloat[]{
\includegraphics[width=0.5\linewidth]{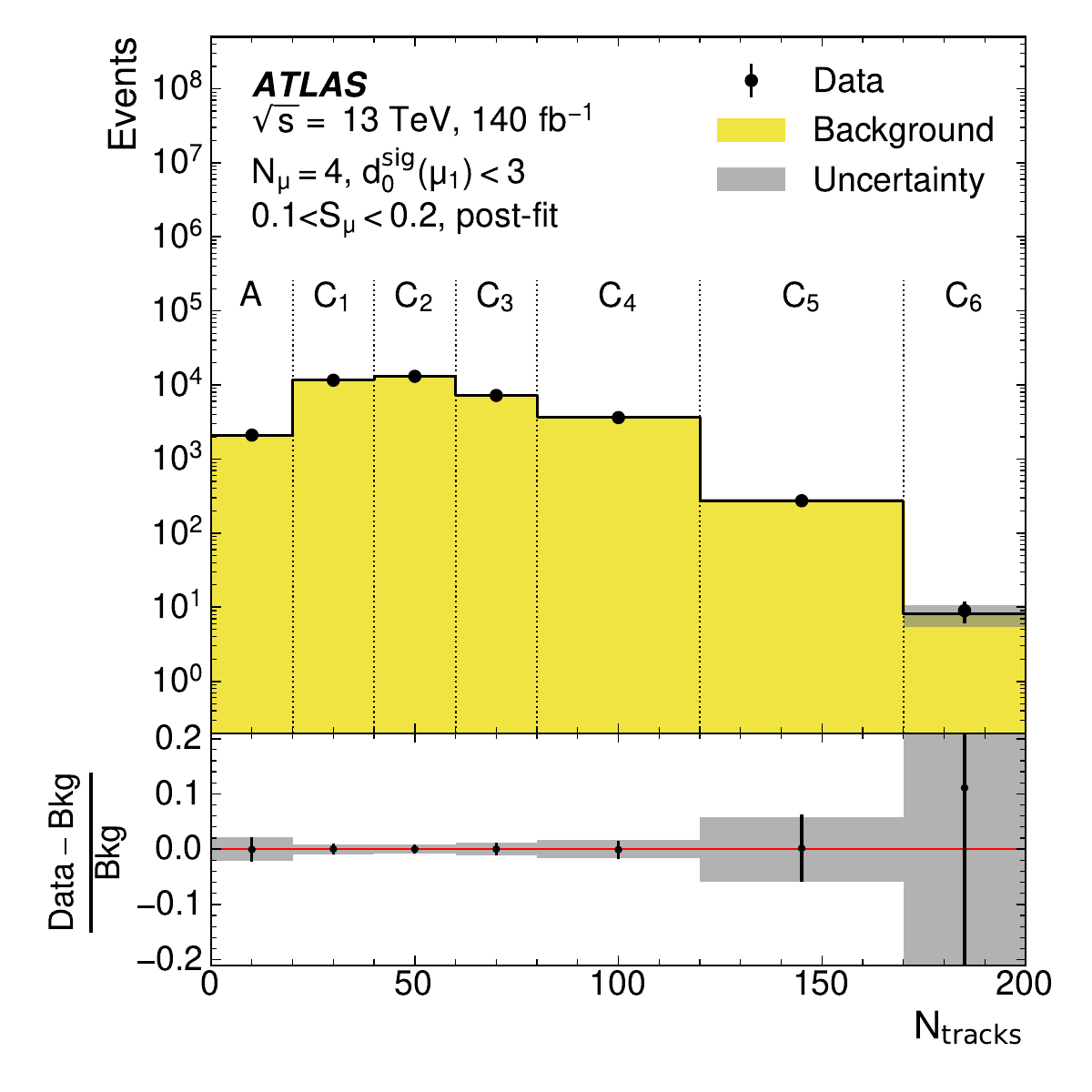}
\vspace{-2mm}
\label{figure_validation_region_a}
\vspace{-2mm}
}
\subfloat[]{
\includegraphics[width=0.5\linewidth]{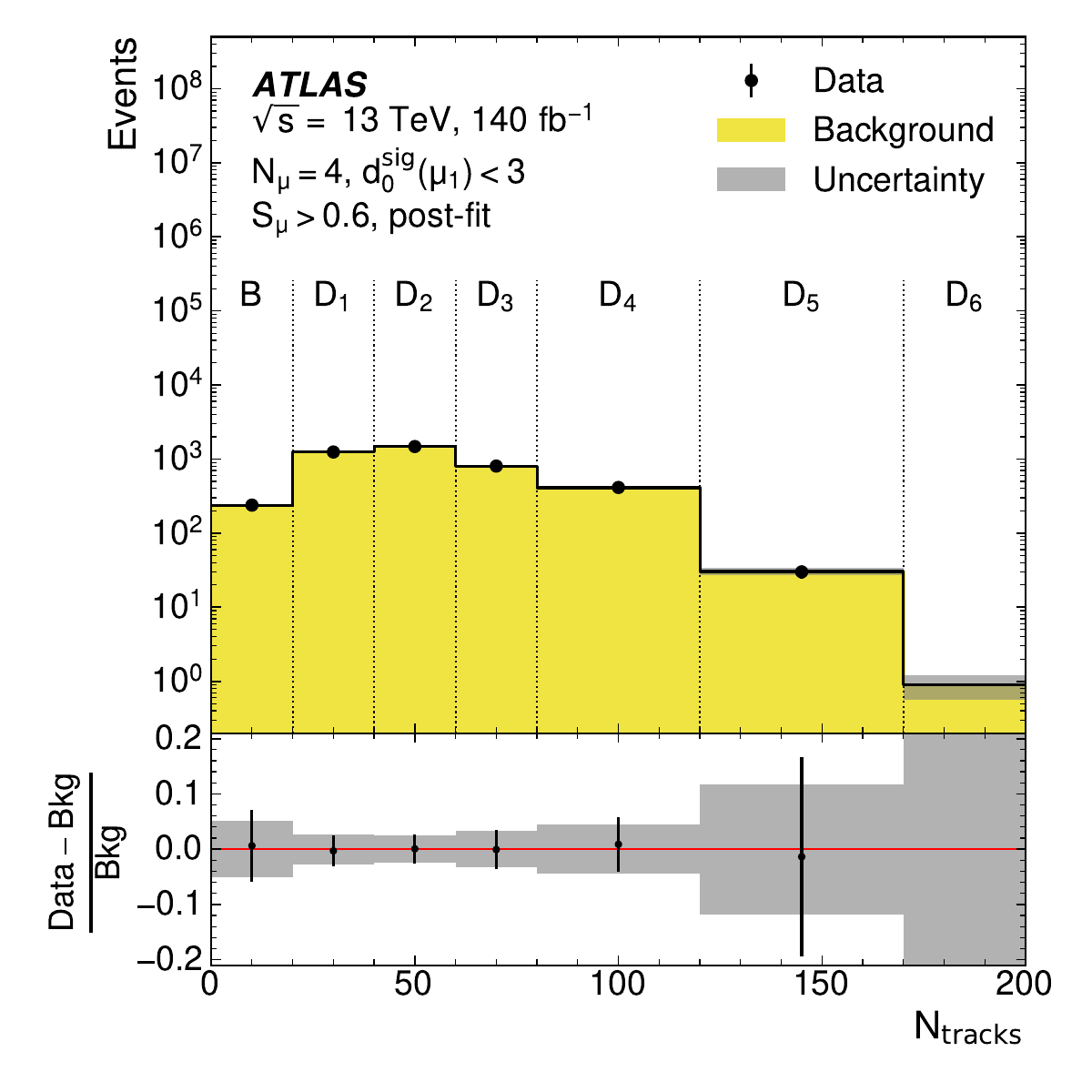}
\vspace{-2mm}
\label{figure_validation_region_b}
\vspace{-2mm}
}\\
\subfloat[]{
\includegraphics[width=0.5\linewidth]{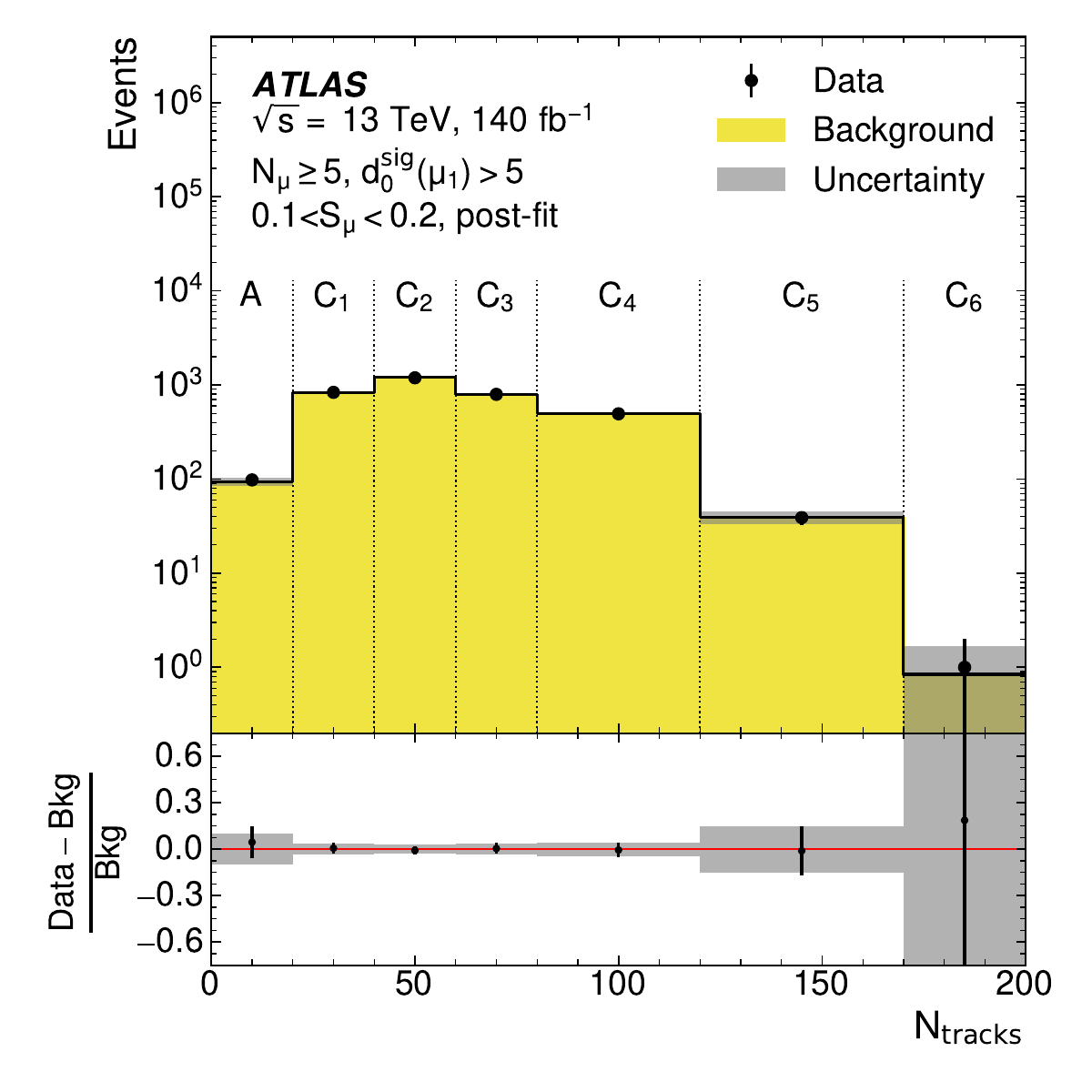}
\vspace{-2mm}
\label{figure_validation_region_c}
\vspace{-2mm}
}
\subfloat[]{
\includegraphics[width=0.5\linewidth]{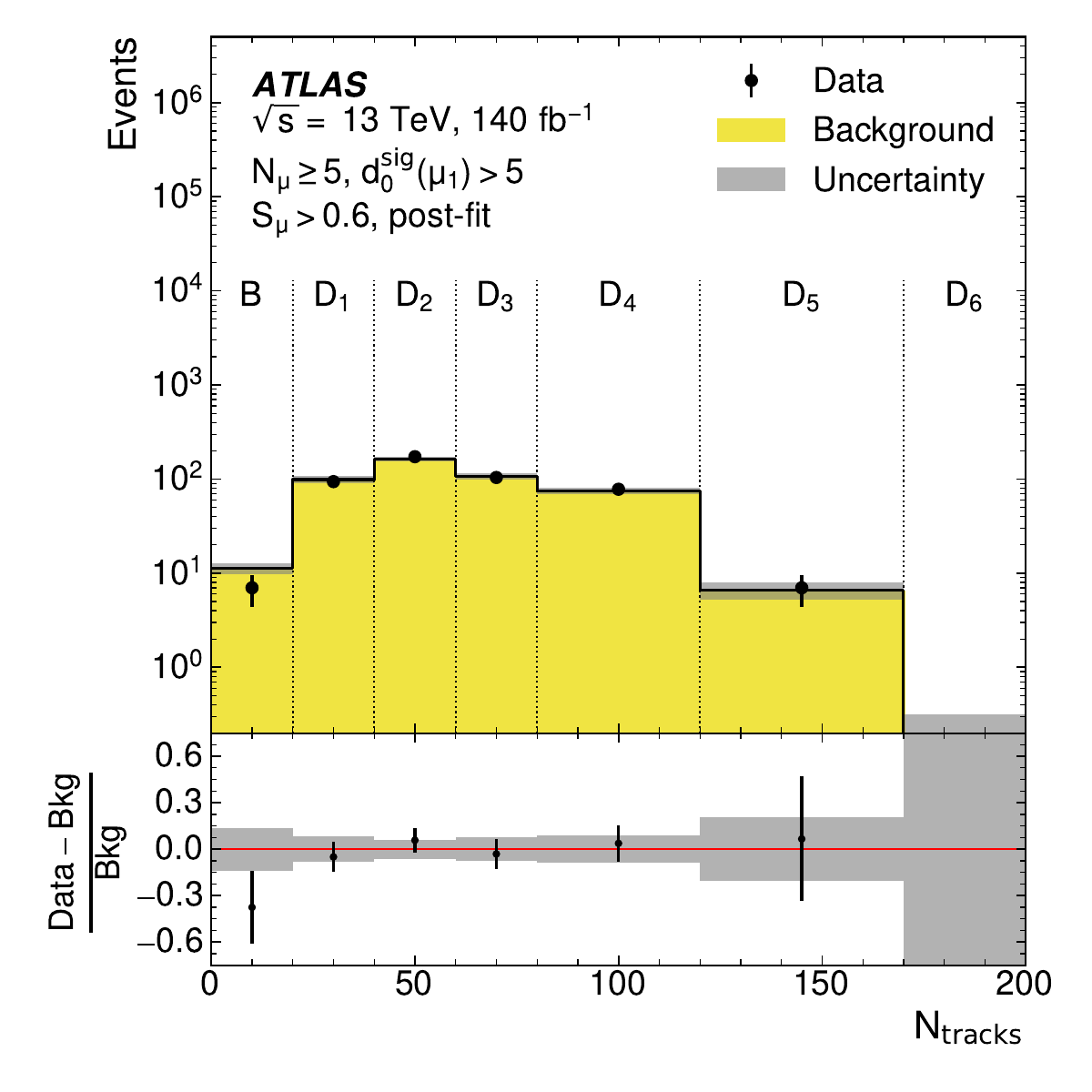}
\vspace{-2mm}
\label{figure_validation_region_d}
\vspace{-2mm}
}
\caption{
Background-only fits in the validation regions with requirements
(\protect\subref{figure_validation_region_a},\protect\subref{figure_validation_region_b}) \(\dsig < 3\) and \(\nmu = 4,\) and
(\protect\subref{figure_validation_region_c},\protect\subref{figure_validation_region_d}) \(\dsig > 5\) and \(\nmu \geq 5,\).
The \(\ntracks\) distributions are shown for
(\protect\subref{figure_validation_region_a},\protect\subref{figure_validation_region_c}) \(0.1 < \sph < 0.2\) and
(\protect\subref{figure_validation_region_b},\protect\subref{figure_validation_region_d}) \(\sph > 0.6\),
with underflow and overflow entries included in the first and last bins, respectively.
The background prediction (post-fit) includes the systematic uncertainties
related to the estimation method described in Section~\ref{sec:sys}.}
\label{figure_validation_region}
\end{figure}

Figure~\ref{figure_validation_region} shows the comparison between data and the background model obtained from background-only fits performed in two representative validation regions.
The goodness-of-fit \(p\)-values for the background-only fits range from 0.48 (for \(\nmu \geq 5,\,\dsig > 5\)) to 0.92, indicating good agreement between data and the background model.

\begin{figure}
\centering
\subfloat[]{
\includegraphics[width=0.5\linewidth]{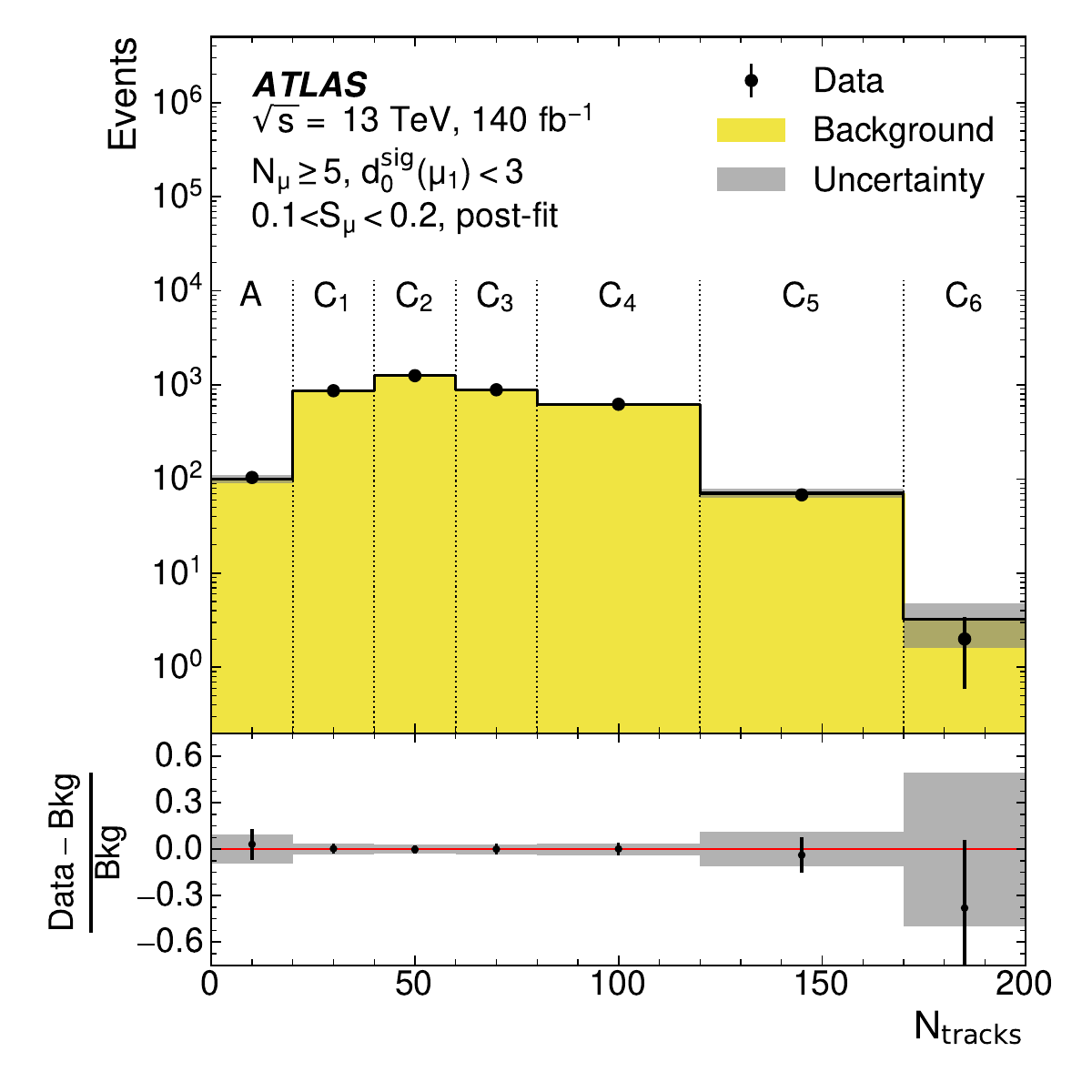}
\vspace{-2mm}
\label{figure_background_only_a}
\vspace{-2mm}
}
\subfloat[]{
\includegraphics[width=0.5\linewidth]{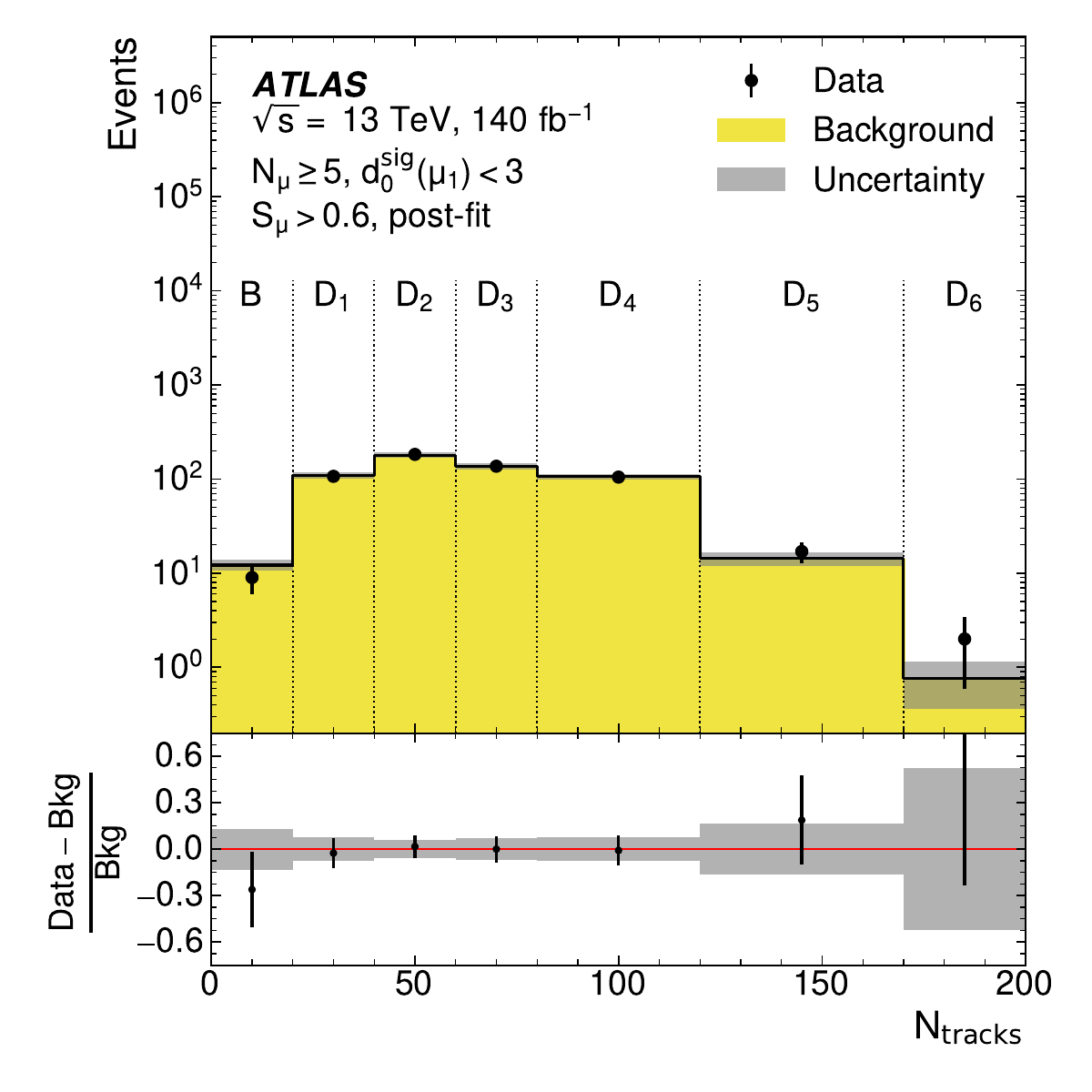}
\vspace{-2mm}
\label{figure_background_only_b}
\vspace{-2mm}
}
\caption{
Background-only fit in the search region defined by \(\nmu \geq 5\) and \(\dsig < 3\).
The \(\ntracks\) distributions are shown for \protect\subref{figure_background_only_a} \(0.1 < \sph < 0.2\)
and \protect\subref{figure_background_only_b} \(\sph > 0.6\), with underflow and overflow entries included
in the first and last bins, respectively.
The background prediction (post-fit) includes the systematic uncertainties
related to the estimation method described in Section~\ref{sec:sys}.
}
\label{figure_background_only}
\end{figure}

\begin{figure}
\centering
\subfloat[]{
\includegraphics[width=0.5\linewidth]{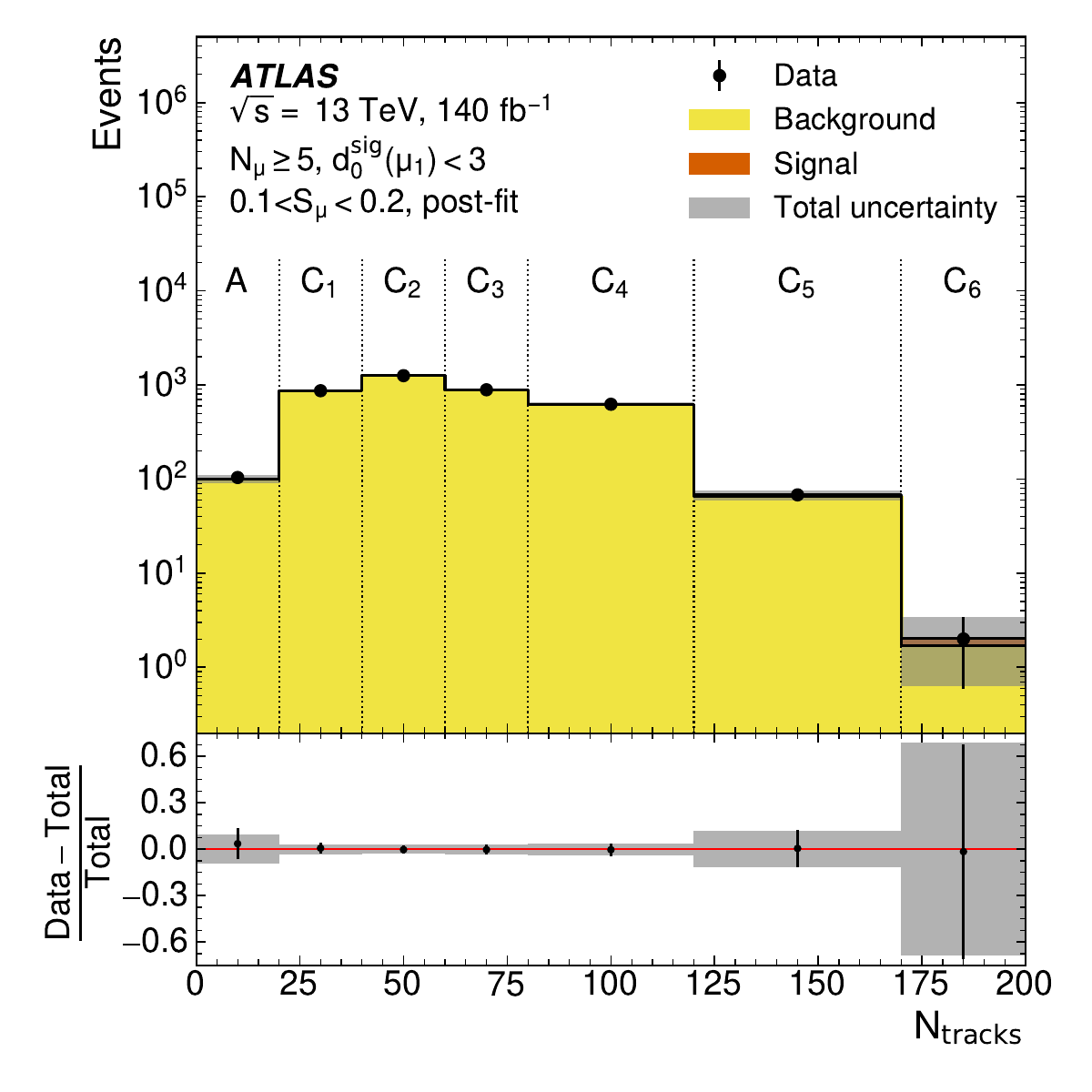}
\vspace{-2mm}
\label{figure_fit_signal_a}
\vspace{-2mm}
}
\subfloat[]{
\includegraphics[width=0.5\linewidth]{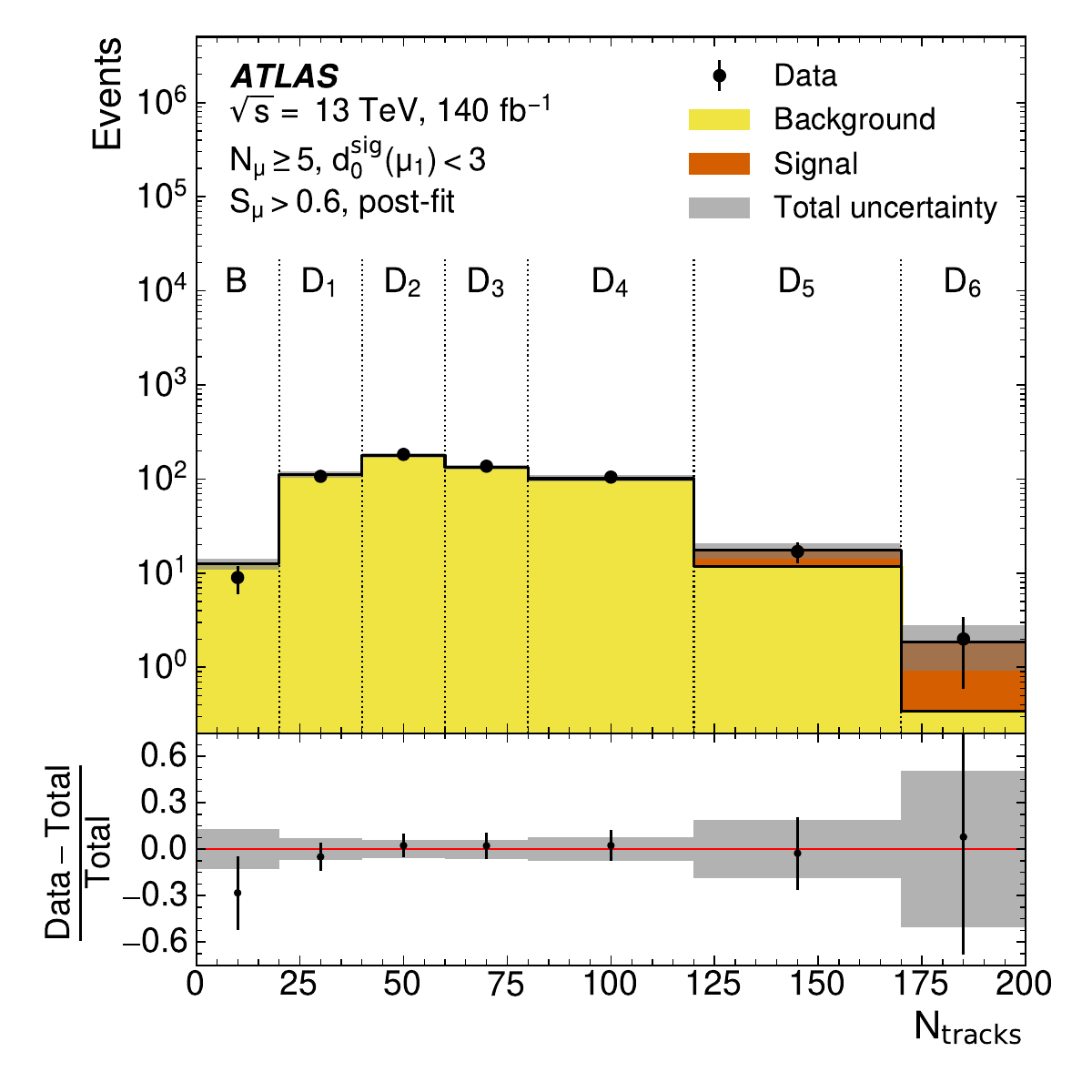}
\vspace{-2mm}
\label{figure_fit_signal_b}
\vspace{-2mm}
}
\caption{
Signal-plus-background fit in the search region defined by \(\nmu \geq 5\) and \(\dsig < 3\) for the signal model
with \(m_S= 400\)~GeV, \(\mAprime = 0.7\)~GeV, \(\mphi = 3\)~GeV, and \(T = 3\)~GeV.
The \(\ntracks\) distributions are shown for \protect\subref{figure_fit_signal_a} \(0.1 < \sph < 0.2\)
and \protect\subref{figure_fit_signal_b} \(\sph > 0.6\),
with underflow and overflow entries included in the first and last bins, respectively.
The background prediction (post-fit) includes the systematic uncertainties
related to the estimation method described in Section~\ref{sec:sys}.
The lower panels show the relative difference between data and the total fitted expectation (signal plus background).
The fitted $\xsec \times \brSUEP$ is $1.64$~fb.
}
\label{figure_fit_signal}
\end{figure}

The distributions observed in data and the background predictions obtained from background-only fits in the search region are shown in Figure~\ref{figure_background_only}.
While a mild excess of events in the highest \(\ntracks\) bin is observed, the fit yields a goodness-of-fit \(p\)-value of 0.42,
indicating good overall agreement between the background-only hypothesis and the data.
The two observed events in the region \(D_6\) have \((\ntracks,\,\nmu,\,\sph) = (188,\,5,\,0.63)\) and \((173,\,7,\,0.68)\), respectively.
The result of a signal-plus-background fit in the search region is shown in Figure~\ref{figure_fit_signal}.
The fit includes the signal model with \(m_S= 400\)~GeV, \(\mAprime = 0.7\)~GeV, \(\mphi = 3\)~GeV, and \(T = 3\)~GeV, which can partially accommodate the excess,
leading to a local significance of \(1.7\sigma\), the largest observed among all tested signal models.

\begin{figure}
\centering
\subfloat[]{
\includegraphics[width=0.5\linewidth]{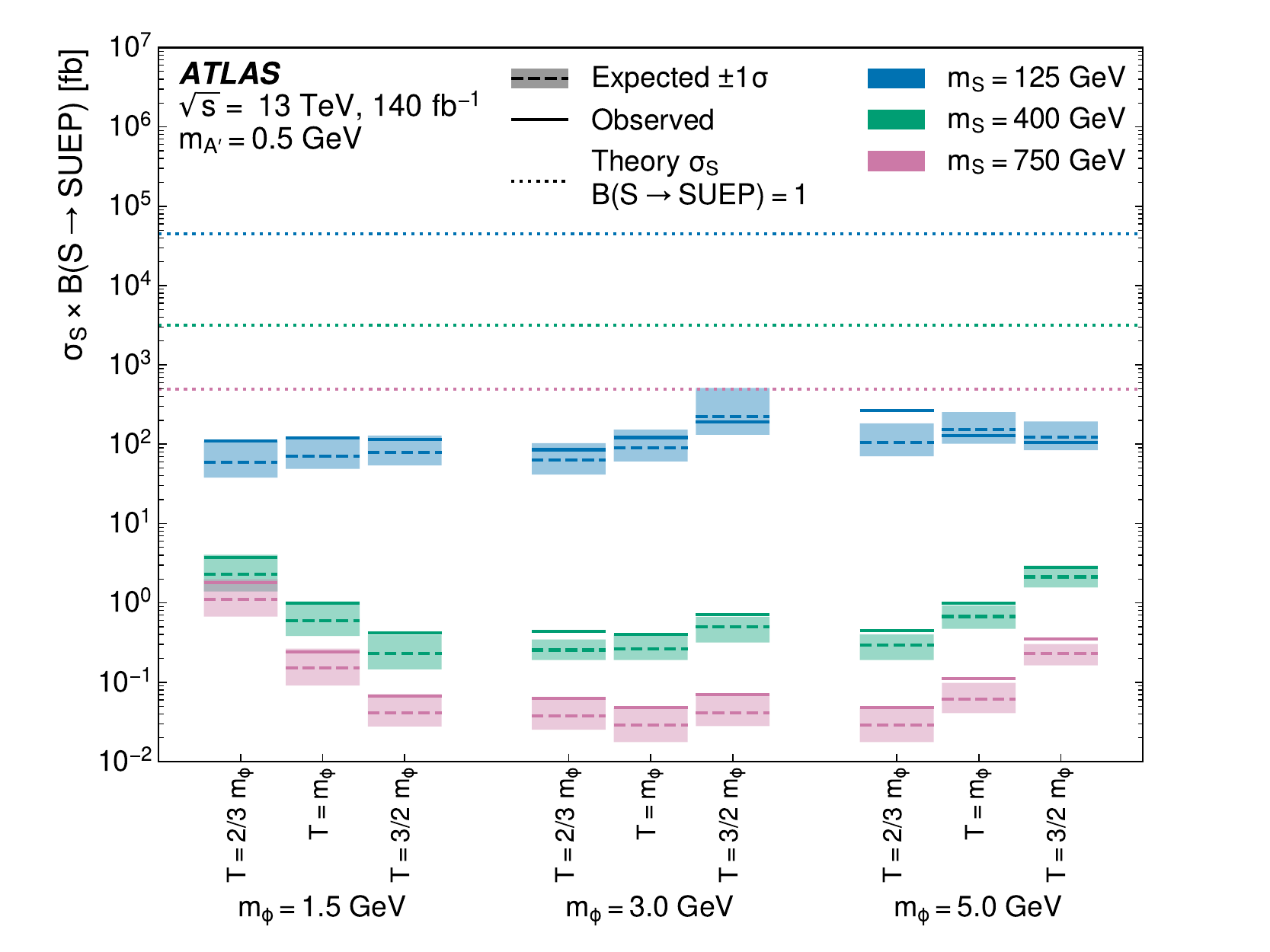}
\label{fig_limits_a}
\vspace{-2mm}
}\\
\subfloat[]{
\includegraphics[width=0.5\linewidth]{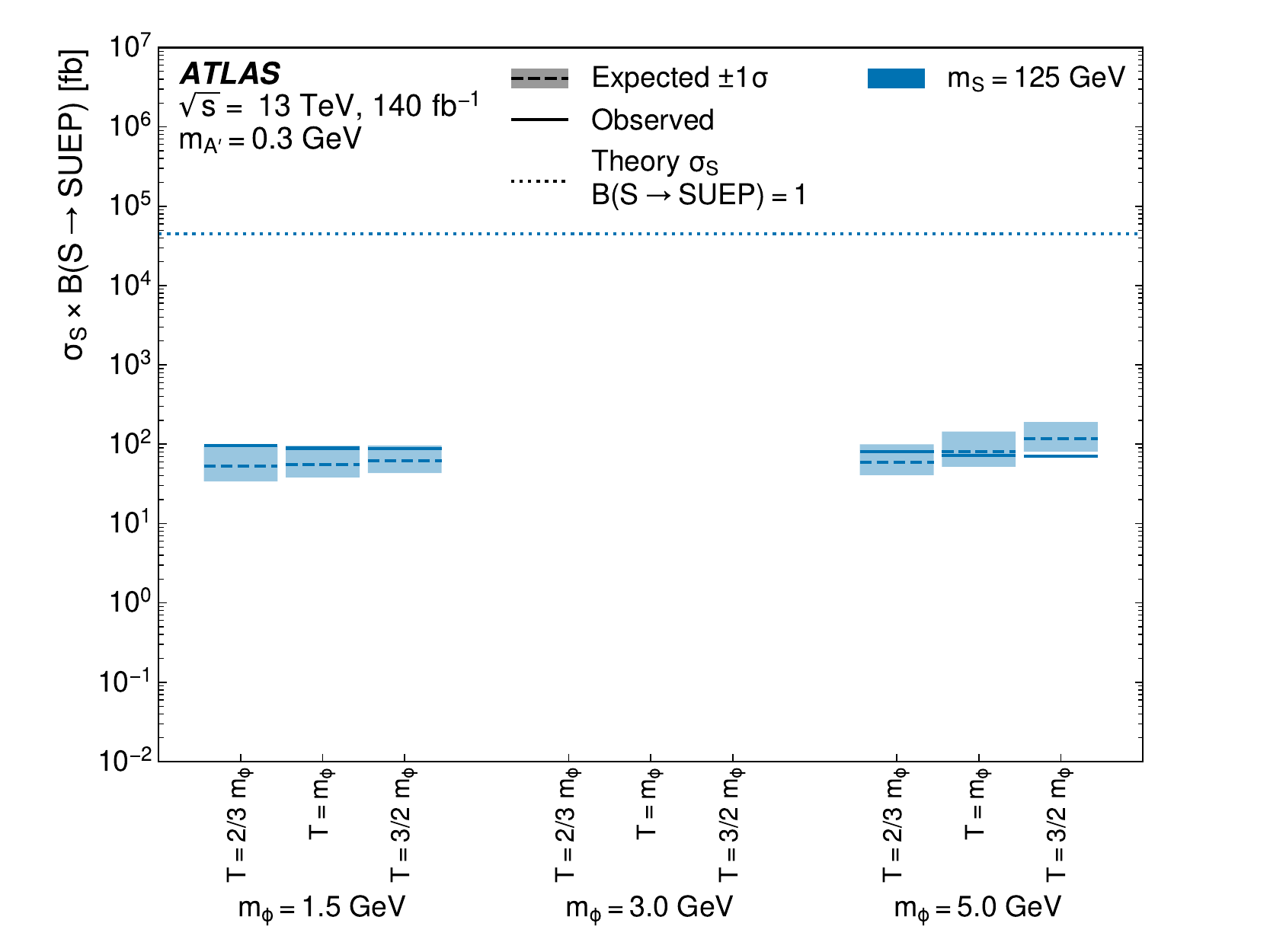}
\label{fig_limits_b}
\vspace{-2mm}
}
\subfloat[]{
\includegraphics[width=0.5\linewidth]{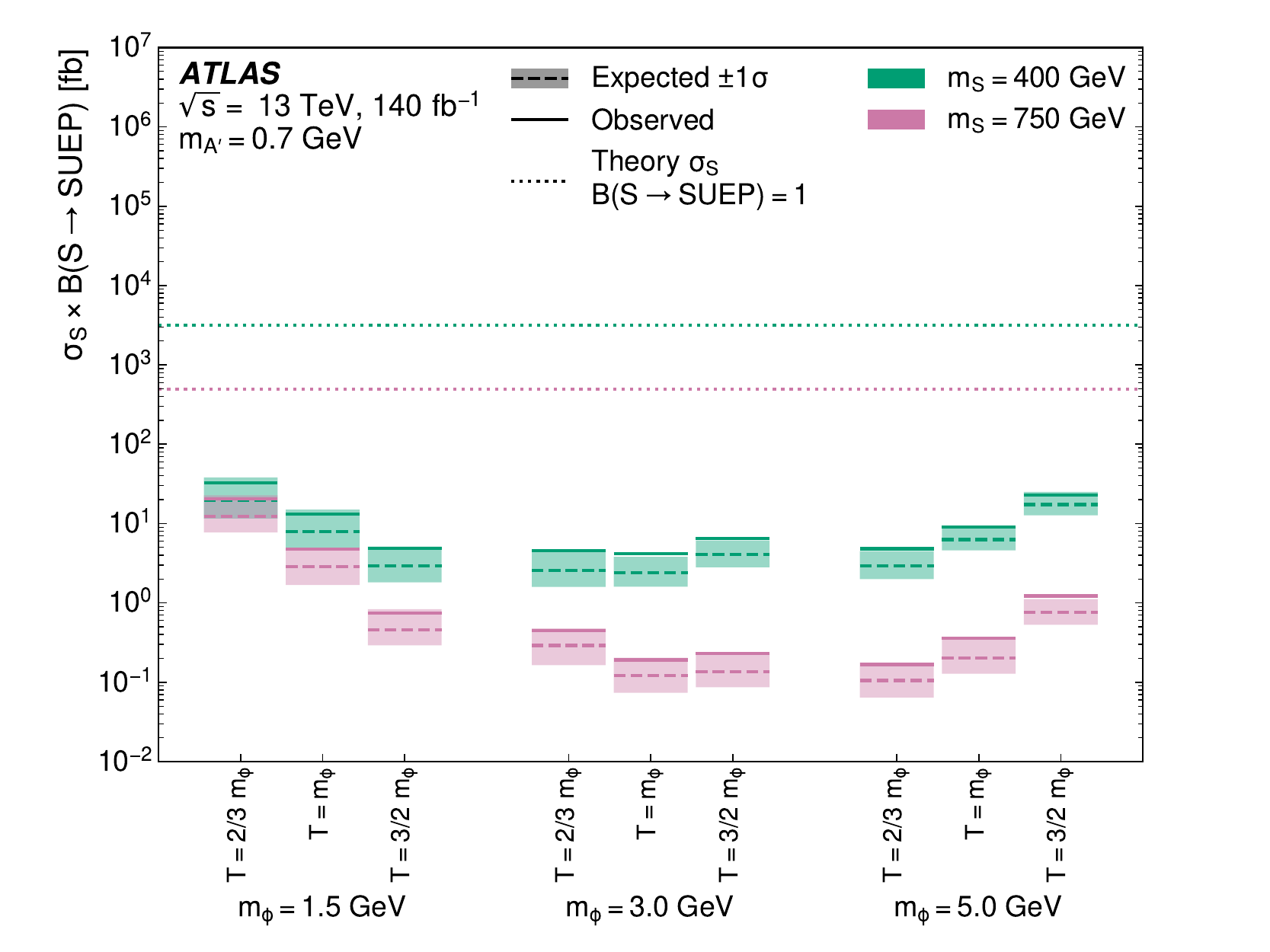}
\label{fig_limits_c}
\vspace{-2mm}
}
\caption{
Expected and observed 95\% CL upper limits on $\xsec \times \brSUEP$ for models with
\protect\subref{fig_limits_a} \(\mAprime = 0.5\)~GeV, \protect\subref{fig_limits_b} \(\mAprime = 0.3\)~GeV,
and \protect\subref{fig_limits_c} \(\mAprime = 0.7\)~GeV.
The theoretical predictions correspond to SM-like gluon-fusion production, as given in Ref.~\cite{LHCHiggsCrossSectionWorkingGroup:2016ypw}.
Missing points in \protect\subref{fig_limits_b} and \protect\subref{fig_limits_c} correspond to signal models that were not tested.
}
\label{fig_limits}
\end{figure}

Given that no significant deviation from the background-only hypothesis is observed, upper limits at the 95\% confidence level (CL) are set on the product of the mediator production cross section and the branching fraction for its decay into a SUEP,
$\xsec \times \brSUEP$, using the \(\mathrm{CL_s}\) method~\cite{Read:2002hq}. The results are shown in Figure~\ref{fig_limits}.
Because of the mild excess observed in the highest \(\ntracks\) bin, the observed limits are slightly weaker than the expected limits, except for signal models predicting lower \(\ntracks\) values.

The limits reach values as low as approximately $0.05$~fb for $m_S = 750$~GeV, $0.4$~fb for $m_S = 400$~GeV, and $70$~fb for $m_S = 125$~GeV.
Similarly, these limits can be interpreted as bounds on $\brSUEP$.
In the case of $m_S = 125$~GeV, this corresponds to an upper limit on the Higgs boson branching fraction of around $0.2\%$, which is well below the limits on the invisible branching fraction obtained in model-independent searches~\cite{ATLAS:2023tkt}.

The strongest limits are generally obtained for $\mphi=T= 3$~GeV, with reduced sensitivity at both lower and higher values of $\mphi$ and $T$.
For $\mphi =T\approx 3$~GeV, the trigger efficiency is high while the separation between signal and background in $\ntracks$ remains strong.
At lower values, the efficiency is reduced due to softer final-state particles, while at higher values the particle multiplicity decreases, leading to reduced discrimination power.

The sensitivity is higher for $\mAprime = 0.5$~GeV, which corresponds to branching fractions
$B(A'\rightarrow e^+e^-,\mu^+\mu^-,\pi^+\pi^-) = (0.4,\,0.4,\,0.2)$,
than for $\mAprime = 0.7$~GeV, where the muon branching fraction is reduced to $0.15$.
The improvement for $\mAprime = 0.3$~GeV, with a muon branching fraction of $0.5$, is less pronounced.

Taking $m_\phi=T=3$~GeV and $m_{A'}=0.5$~GeV as an example, the observed upper limits on the mediator production cross section via gluon--gluon fusion for the probed mediator masses
are between one and four orders of magnitude more stringent than previous results~\cite{CMS:2024nca}.
Likewise, assuming the 125~GeV mediator is the SM Higgs boson, the upper limit on $B(H \to \mathrm{SUEP})$ is around $0.3\%$ for $m_\phi=T=3,5$~GeV and $m_{A'}=0.5$~GeV,
which significantly improves upon those reported for similar model parameters in Ref.~\cite{CMS:2024nca,CMS:2026jia}.

\newpage

\FloatBarrier

\section{Conclusion}
\label{sec:conclusion}

A search for soft unclustered energy patterns is presented using \(140~\mathrm{fb}^{-1}\) of \(\sqrt{s} = 13\)~TeV proton--proton collision data recorded by the ATLAS detector.
The analysis targets scenarios in which a scalar mediator decays into a strongly interacting, quasi-conformal hidden sector, giving rise to high-multiplicity final states containing low-momentum particles, including muons,
that are distributed approximately isotropically in the mediator center-of-mass frame.

Events are selected using multi-muon triggers, and the search region is defined by requirements on the muon multiplicity, average muon transverse momentum, and transverse impact-parameter significance, targeting prompt, high-multiplicity final states with soft muons.
The discrimination between signal and the dominant QCD multijet background is achieved using two observables:
the muon sphericity, constructed in the rest frame of the muon system, and the charged-particle track multiplicity, corrected for pile-up.
The background contribution is estimated by using a likelihood-based ABCD method that exploits the weak correlation between these observables, with residual correlations accounted for through a data-driven parameterization.

No significant deviation from the background-only hypothesis is observed in the search region.
The results are therefore interpreted in terms of upper limits at the 95\% confidence level on the on the product of the mediator production cross section and the branching fraction for its decay into a SUEP.

This search extends the sensitivity to SUEP scenarios beyond previous LHC results~\cite{CMS:2024nca,CMS:2026jia}, particularly in regions of parameter space characterized by
moderate particle multiplicities and sizable muon fractions in the final state.
The analysis demonstrates the effectiveness of multi-muon-based strategies and provides a complementary approach
to existing searches targeting primarily hadronic final states.

\section*{Acknowledgements}
%

%

% The next lines are included from the .//acknowledgements/Acknowledgements.tex input file
%
%

%
%

We thank CERN for the very successful operation of the LHC and its injectors, as well as the support staff at
CERN and at our institutions worldwide without whom ATLAS could not be operated efficiently.

The crucial computing support from all WLCG partners is acknowledged gratefully, in particular from CERN, the ATLAS Tier-1 facilities at TRIUMF/SFU (Canada), NDGF (Denmark, Norway, Sweden), CC-IN2P3 (France), KIT/GridKA (Germany), INFN-CNAF (Italy), NL-T1 (Netherlands), PIC (Spain), RAL (UK) and BNL (USA), the Tier-2 facilities worldwide and large non-WLCG resource providers. Major contributors of computing resources are listed in Ref.~\cite{ATL-SOFT-PUB-2026-001}.

We gratefully acknowledge the support of ANPCyT, Argentina; YerPhI, Armenia; ARC, Australia; BMWFW and FWF, Austria; ANAS, Azerbaijan; CNPq and FAPESP, Brazil; NSERC, NRC and CFI, Canada; CERN; ANID, Chile; CAS, MOST and NSFC, China; Minciencias, Colombia; MEYS CR, Czech Republic; DNRF and DNSRC, Denmark; IN2P3-CNRS and CEA-DRF/IRFU, France; SRNSFG, Georgia; BMFTR, HGF and MPG, Germany; GSRI, Greece; RGC and Hong Kong SAR, China; ICHEP and Academy of Sciences and Humanities, Israel; INFN, Italy; MEXT and JSPS, Japan; CNRST, Morocco; NWO, Netherlands; RCN, Norway; MNiSW, Poland; FCT, Portugal; MNE/IFA, Romania; MSTDI, Serbia; MSSR, Slovakia; ARIS and MVZI, Slovenia; DSI/NRF, South Africa; MICIU/AEI, Spain; SRC and Wallenberg Foundation, Sweden; SERI, SNSF and Cantons of Bern and Geneva, Switzerland; NSTC, Taipei; TENMAK, T\"urkiye; STFC/UKRI, United Kingdom; DOE and NSF, United States of America.

Individual groups and members have received support from BCKDF, CANARIE, CRC and DRAC, Canada; CERN-CZ, FORTE and PRIMUS, Czech Republic; COST, ERC, ERDF, Horizon 2020 and Marie Sk{\l}odowska-Curie Actions, European Union; Investissements d'Avenir Labex, Investissements d'Avenir Idex and ANR, France; DFG and AvH Foundation, Germany; Herakleitos, Thales and Aristeia programmes co-financed by EU-ESF and the Greek NSRF, Greece; BSF-NSF and MINERVA, Israel; NCN and NAWA, Poland; La Caixa Banking Foundation, CERCA and AGAUR programs from Generalitat de Catalunya and PROMETEO and GenT Programmes Generalitat Valenciana, Spain; G\"{o}ran Gustafssons Stiftelse, Sweden; The Royal Society and Leverhulme Trust, United Kingdom; Eric and Wendy Schmidt Fund for Strategic Innovation, United States of America.

In addition, individual members wish to acknowledge support from Chile: Agencia Nacional de Investigaci\'on y Desarrollo (ANID FONDECYT reg. 1230987, FONDECYT 1230812, FONDECYT 1240864, Fondecyt 3240661, Fondecyt Regular 1240721); China: Chinese Ministry of Science and Technology (MOST-2023YFA1605700, MOST-2023YFA1609300), National Natural Science Foundation of China (NSFC 12275265, NSFC-W2543005); Czech Republic: Czech Science Foundation (GACR - 24-11373S), Ministry of Education Youth and Sports (ERC-CZ-LL2327, FORTE CZ.02.01.01/00/22\_008/0004632); EU: H2020 European Research Council (ERC - 101002463); European Union: European Research Council (BARD No. 101116429, ERC - 101219398, ERC 101089007), European Regional Development Fund (HE COFUND GA No.101081355, ERDF), Marie Sklodowska-Curie Actions (GAP-101168829); France: Agence Nationale de la Recherche (ANR-21-CE31-0013, ANR-22-EDIR-0002, ANR-24-CE31-0504-01); Germany: Deutsche Forschungsgemeinschaft (DFG - 469666862); China: Research Grants Council (GRF); Italy: Istituto Nazionale di Fisica Nucleare (LHC-MIUR - 28003/2025), Ministero dell'Università e della Ricerca (NextGenEU I53D23000820006 M4C2.1.1, SOE2024\_0000023); Japan: Japan Society for the Promotion of Science (JSPS KAKENHI  JP25H0063, JSPS KAKENHI JP22H04944, JSPS KAKENHI JP24K23939, JSPS KAKENHI JP24KK0251, JSPS KAKENHI JP25H00650, JSPS KAKENHI JP25H01291, JSPS KAKENHI JP25K01011, JSPS KAKENHI JP25K01023, JSPS KAKENHI JP25KK0047); Poland: Polish National Science Centre (NCN 2021/42/E/ST2/00350, NCN OPUS 2023/51/B/ST2/02507, NCN OPUS nr 2022/47/B/ST2/03059, UMO-2022/47/O/ST2/00148, UMO-2023/49/B/ST2/04085, UMO-2023/51/B/ST2/00920, UMO-2024/53/N/ST2/00869); Spain: Agència de Gestió d'Ajuts Universitaris i de Recerca. (AGAUR - 2023 BP 00141), Ministry of Science and Innovation (RYC2019-028510-I, RYC2020-030254-I, RYC2021-031273-I, RYC2022-038164-I), Ministerio de Ciencia, Innovación y Universidades/Agencia Estatal de Investigaci\'on (EU NextGenerationEU (PRTR-C17.I1), PID2022-142604OB-C22); Sweden: Carl Trygger Foundation (Carl Trygger Foundation CTS 22:2312), Swedish Research Council (Swedish Research Council 2023-04654, VR 2021-03651, VR 2022-03845, VR 2022-04683, VR 2023-03403, VR 2024-05451, VR 2025-05940), Knut and Alice Wallenberg Foundation (KAW 2023.0366); United Kingdom: The Binks Trust, Royal Society (NIF-R1-231091); United States of America: U.S. Department of Energy (ECA DE-AC02-76SF00515), John Templeton Foundation (John Templeton Foundation 63206), Neubauer Family Foundation.

%
%

% End of text imported from the .//acknowledgements/Acknowledgements.tex input file

%
%
%
%
%
%

%
%

\clearpage
\printbibliography
\clearpage
% ATLAS Collaboration author list
% Reference date of EXOT-2018-56 is 2026-04-17
% Author list last updated on date 19-MAY-26
% Data extracted on 19-May-2026 for paper reference EXOT-2018-56
% at 4:48pm
 
\begin{flushleft}
\hypersetup{urlcolor=black}
{\Large The ATLAS Collaboration}

\bigskip

\AtlasOrcid[0000-0002-6665-4934]{G.~Aad}$^\textrm{\scriptsize 102}$,
\AtlasOrcid[0000-0001-7616-1554]{E.~Aakvaag}$^\textrm{\scriptsize 17}$,
\AtlasOrcid[0000-0002-5888-2734]{B.~Abbott}$^\textrm{\scriptsize 121}$,
\AtlasOrcid[0000-0002-0287-5869]{S.~Abdelhameed}$^\textrm{\scriptsize 83b}$,
\AtlasOrcid[0000-0002-1002-1652]{K.~Abeling}$^\textrm{\scriptsize 54}$,
\AtlasOrcid[0000-0001-5763-2760]{N.J.~Abicht}$^\textrm{\scriptsize 48}$,
\AtlasOrcid[0000-0002-8496-9294]{S.H.~Abidi}$^\textrm{\scriptsize 30}$,
\AtlasOrcid[0009-0003-6578-220X]{M.~Aboelela}$^\textrm{\scriptsize 44}$,
\AtlasOrcid[0000-0002-9987-2292]{A.~Aboulhorma}$^\textrm{\scriptsize 36e}$,
\AtlasOrcid[0000-0001-5329-6640]{H.~Abramowicz}$^\textrm{\scriptsize 154}$
% ...
\bigskip
\\

% ...
$^{17}$Department for Physics and Technology, University of Bergen, Bergen; Norway.\\
% ...
$^{30}$Physics Department, Brookhaven National Laboratory, Upton NY; United States of America.\\
% ...
$^{36}$$^{(a)}$Facult\'e des Sciences Ain Chock, Universit\'e Hassan II de Casablanca;$^{(b)}$Facult\'{e} des Sciences, Universit\'{e} Ibn-Tofail, K\'{e}nitra;$^{(c)}$Facult\'e des Sciences Semlalia, Universit\'e Cadi Ayyad, LPHEA-Marrakech;$^{(d)}$LPMR, Facult\'e des Sciences, Universit\'e Mohamed Premier, Oujda;$^{(e)}$Facult\'e des sciences, Universit\'e Mohammed V, Rabat;$^{(f)}$Institute of Applied Physics, Mohammed VI Polytechnic University, Ben Guerir; Morocco.\\
% ...
$^{44}$Physics Department, Southern Methodist University, Dallas TX; United States of America.\\
% ...
$^{48}$Fakult\"{a}t Physik, Technische Universit{\"a}t Dortmund, Dortmund; Germany.\\
% ...
$^{54}$II. Physikalisches Institut, Georg-August-Universit\"{a}t G\"ottingen, G\"ottingen; Germany.\\
% ...
$^{83}$$^{(a)}$Khalifa University of Science and Technology, Abu Dhabi;$^{(b)}$New York University Abu Dhabi, Abu Dhabi;$^{(c)}$United Arab Emirates University, Al Ain;$^{(d)}$University of Sharjah, Sharjah; United Arab Emirates.\\
% ...
$^{102}$CPPM, Aix-Marseille Universit\'e, CNRS/IN2P3, Marseille; France.\\
% ...
$^{121}$Homer L. Dodge Department of Physics and Astronomy, University of Oklahoma, Norman OK; United States of America.\\
% ...
$^{154}$Raymond and Beverly Sackler School of Physics and Astronomy, Tel Aviv University, Tel Aviv; Israel.\\
% ...
$^{a}$ Also at Affiliated with an institute formerly covered by a cooperation agreement with CERN.\\
$^{b}$ Also at An-Najah National University, Nablus; Palestine.\\
$^{c}$ Also at Borough of Manhattan Community College, City University of New York, New York NY; United States of America.\\
$^{d}$ Also at Center for Interdisciplinary Research and Innovation (CIRI-AUTH), Thessaloniki; Greece.\\
$^{e}$ Also at Centre of Physics of the Universities of Minho and Porto (CF-UM-UP); Portugal.\\
$^{f}$ Also at CERN, Geneva; Switzerland.\\
% ...

\end{flushleft}

% Created with Glance <Atlas.Glance@cern.ch>

%

%
%

%

%
%
%
%

\end{document}